\documentclass[prb,twocolumn,showpacs,amsmath,amssymb]
{revtex4}

\usepackage[dvips]{color}
\usepackage{graphicx}

\begin{document}

\title{Finite temperature study of bosons in a two dimensional optical lattice}

\author{K. W. Mahmud,$^1$ E. N. Duchon,$^2$ Y. Kato,$^3$ N. Kawashima,$^4$
R. T. Scalettar,$^1$ and N. Trivedi$^2$}
\affiliation{$^1$Department of Physics, University of California,
Davis, California 95616, USA \\
$^2$Department of Physics, Ohio State University, Columbus, Ohio
43210, USA\\
$^3$Theoretical Division, Los Alamos National Laboratory, Los
Alamos, New Mexico 87545, USA\\
$^4$Institute for Solid State Physics, University of Tokyo,
Kashiwa, Chiba 277-8581, Japan}

\begin{abstract}
We use quantum Monte Carlo (QMC) simulations to study the combined
effects of harmonic confinement and temperature for bosons in a
two dimensional optical lattice. The scale invariant, finite
temperature, state diagram is presented for the Bose-Hubbard model
in terms of experimental parameters -- the particle number,
confining potential and interaction strength. To distinguish the
nature of the spatially separated superfluid, Mott Insulator and
normal Bose liquid phases, we examine the local density,
compressibility, superfluid density and Green's function. In the
annular superfluid rings, as the width of the ring decreases, the
long range superfluid correlations start to deviate from an
equivalent homogeneous 2D system. At zero temperature, the
correlation decay is intermediate between 1D and 2D, while at
finite temperature, the decay is similar to that in 1D at a much
lower temperature. The calculations reveal shortcomings of the
local density approximation (LDA) in describing superfluid
properties of trapped bosons. We also present the finite
temperature phase diagram for the homogeneous two dimensional
Bose-Hubbard model. We compare our state diagram with the results
of a recent experiment at NIST on a harmonically trapped 2D
lattice [Phys. Rev. Lett. {\bf 105}, 110401 (2010)], and identify
a finite temperature effect in the experiment.
\end{abstract}

\pacs{03.75.Hh,03.75.Lm,05.30.Jp,67.85.-d}

\maketitle

%%%%%%%%%

\section{Introduction}

Much progress has been made in the last decade in the use of
trapped, ultracold atoms for the optical lattice emulation of
tight binding Hamiltonians
\cite{greiner02,jaksch98,lewenstein07,bloch08}. In the bosonic
case, Quantum Monte Carlo (QMC) simulations are possible on very
large lattices (algorithms scale linearly with the system size)
and at low temperatures (no sign problem).  As a result, close
contact between experiments and theory has been possible, with
successful quantitative comparisons of momentum distributions and
phase transition critical
points~\cite{spielman10,spielman0708,trotzky09}. Recent QMC
simulations of the homogeneous system have refined early
determinations \cite{batrouni90,trivedi91,freericks96} of the
critical point for the ground state superfluid-Mott insulator
boundary to very high accuracy~\cite{sansone08}, and have also
examined the finite temperature behavior at integer filling
\cite{sansone08}. QMC studies which include a confining potential
and hence determine a ``state diagram" showing which phases
coexist as a function of interaction strength and characteristic
density have also been reported at low temperature \cite{rigol09}
and compared to experiment \cite{spielman10,spielman0708}.

In current experiments,
\cite{spielman10,spielman0708,trotzky09,cchin09,cchin10,bakr09,sherson10}
ultracold atoms are first loaded in a harmonic trap, and then an
external sinusoidal potential is slowly ramped up to create the
optical lattice. By varying the depth of the lattice, the
interaction strength ($U/t$) is changed, thereby driving the
system through a quantum phase transition from a superfluid (SF)
phase to a Mott insulating (MI) phase. Although the initial system
can be prepared at a relatively low temperature, the ensuing
system after ramp-up of the lattice has a temperature which is
usually higher due to adiabatic and other heating mechanisms
\cite{cchin10,pupillo06,pollet08,ho07,mahmud10}. Recent
experiments have reported temperatures on the order of $T/t
\approx 0.9/k_B$~\cite{spielman10}. At such temperatures, the
effects of excited states become important, motivating
investigations into the finite temperature phase diagram, quantum
versus thermal transitions, the location of spatially separated
phases in a trap at finite-T, etc. Aspects of finite-T effects on
the MI have been discussed in
Ref.~\cite{sansone08,gerbier07,demarco07}. Several QMC studies
have also appeared at zero and finite temperatures, dealing with
quantum criticality and phase coexistence in a
trap~\cite{wessel04,zhou09,kato08,roscilde10,pollet10,fang10,hazzard10,ma10}.

The goal in this paper is to provide a study of the combined
effects of a confining potential and finite temperature on the
state diagram of the Bose-Hubbard model in two dimensions,
generalizing previous finite-T QMC work \cite{sansone08} at fixed
density, and ground state QMC studies in the presence of a
confining potential \cite{batrouni02,rigol09}. In addition to the
trapped state diagram, we present the finite temperature phase
diagram of the homogeneous 2D Bose-Hubbard model. Our results
quantify the spatial inhomogeneity/phase coexistence which is
created by the trap and the interplay of thermal fluctuations,
which give rise to a third, normal (N) liquid phase in addition to
the usual ground state SF and MI phases of the Bose Hubbard model
\cite{fisher89}. We use an appropriately scaled measure of
particle number, the `characteristic density', which allows
systems of different sizes to be compared \cite{rigol09}, to
present the scale invariant finite temperature state diagram. We
compare our state diagrams with the results of a recent NIST
experiment on a harmonically trapped 2D lattice~\cite{spielman10},
and identify a finite temperature effect in the experimental data.
To better understand the nature of the trapped phases, we
investigate the correlation function decay. In the annular
superfluid rings, the correlation decay is different from an
equivalent homogeneous 2D superfluid -- matching for a short
distance and then falling off at a faster rate for longer
distances. At zero temperature, the correlation decay is
intermediate between 1D and 2D, while at finite temperature, the
decay is similar to 1D at a much lower temperature. In short, the
width of the trap and the temperature determine the 2D to 1D
crossover. These findings provide evidence for the breakdown of
the local density approximation (LDA) for the description of
superfluid properties of trapped bosons.

The article is organized as follows. In Sec.~II, the Bose-Hubbard
model~\cite{fisher89} and the observables used to monitor the
state diagram are introduced. The QMC methodology is also briefly
described. In Sec.~III, we present the QMC homogeneous system
phase diagram at finite-T. In Sec. IV, we analyze finite
temperature effects in a harmonically trapped system, present the
state diagram at zero and finite temperature, and compare them
with a recent NIST experiment~\cite{spielman10}. In Sec. V, we
examine the dependence of spatial correlations in the trapped
superfluid rings, and present evidence for the breakdown of LDA.
Finally, we summarize our results in Sec.~VI.

\section{Model and computational methods}

Cold bosonic atoms in the lowest band of an optical lattice can be
modeled by the Bose-Hubbard model,
\begin{eqnarray}
H &=& -t \sum_{\langle i,j \rangle} \left( a^\dagger_{i} a^{}_{j}
+ a^\dagger_{j} a^{}_{i} \right)
+ \dfrac{U}{2} \sum_i n_{i} \left( n_{i} -1 \right) \nonumber \\
&&+ V_T \sum_{i} r_i^2\ n_{i}- \mu \sum_{i} n_{i} \,\,,
\label{HubbB}
\end{eqnarray}
Here $a^\dagger_{i}$, $a_{i}$ are the creation and annihilation
operators of a boson at site $i$, $n_{i} = a^\dagger_{i} a^{}_{i}$
is the number of bosons, $t$ is the strength of hopping
\cite{foot1} between nearest neighbors (here on a square lattice),
and $U$ is the on-site repulsive interaction. $V_T$ is the
curvature of the external harmonic trap which introduces
inhomogeneity to the system. $r_i^2= x_i^2+y_i^2$ is the
distance from the center of the trap, $x_i=d \times i$, where $d$
is the period of the lattice, and $i$ an integer. The hopping $t$,
interaction $U$, and trap curvature $V_T$ are tunable parameters
in the experiment.

In this paper we use the modified directed-loop algorithm
\cite{kato0709} for the QMC simulation based on the Feynman path
integral \cite{kawashima04}, since the simple application of
directed-loop algorithm to the Bose-Hubbard model, that has the
widest applicability, is not efficient especially for large $U/t$.

To obtain the phase diagram, we examine three local observables --
the density $\rho_i=\langle \,n_i \,\rangle$, compressibility
$\kappa_i$, superfluid density $\rho_s$, and Green's function
$g(i,j)$. Together, these variables are able to distinguish the
spatially separated phases in a trap.  The local compressibility
is a measure of the response of the local density to a change in
the chemical potential,
\begin{equation}
\kappa_i =\frac{\partial \langle n_i \rangle}{\partial \mu_i}=
\int_0^{\beta} d\tau[\langle n_i(\tau) n_i(0) \rangle-\langle
n_i(\tau) \rangle \langle n_i(0) \rangle] \label{nk}
\end{equation}
Here $\mu_i=\mu-V_T r_i^2$ is the site dependent local chemical
potential which decreases away from the trap center, and
$\beta=1/T$ is the inverse temperature. For a trapped system, a
qualitative criterion for the appearance of a local MI is that the
local density be an integer, as for the homogeneous case. At
$T=0$, a more precise criterion for identifying a local MI is to
require that $\kappa_i$ take on a value equal to the
compressibility, $\kappa=\partial \rho/\partial \mu$, of a
homogeneous system of commensurate
filling~\cite{batrouni02,rigol09}. It is typically the case that
an extended (more than 4-5 sites) region of nearly constant,
integer density has $\kappa_i$ close to its value in the
homogeneous MI, thus providing a more rapid, visual means to
identify a likely MI region. For finite-T, since the MI-N boundary
is a crossover, there is no sharply defined phase boundary, and
there is arbitrariness in drawing such a boundary. In delineating
the crossover boundary for the homogeneous and the trapped system
at finite temperature, we have taken $\kappa < 0.02/U$ to identify
the Mott region. A superfluid region has a nonzero superfluid
density, while the normal region is identified as the region with
zero superfluid density and incommensurate filling with $\kappa >
0.02/U$.

The Green's function, or one-particle density matrix,
\begin{equation}
g(i,j)=\langle a^\dagger_{i} a_{j} \rangle
\end{equation}
is useful in identifying SF regions.  Specifically, a $g(i,j)$
which falls off sufficiently slowly as $|i-j|$ becomes large
signals the development of off-diagonal long range order.
Conversely, a rapid (exponential) fall-off of $g(i,j)$ is
consistent with the appearance of a N or MI phase.

To define a local SF density is more subtle, since $\rho_s$
involves the response of the system to a global phase twist. While
there may be a way to generalize this to a local quantity, here we
adopt the simpler LDA approach, assigning the homogeneous system
value of $\rho_s$ to different spatial locations of the confined
system~\cite{zhou09}. This gives an initial state diagram for
confined systems. The correlation function $g(i,j)$ provides more
information about the superfluid properties of the finite width
annular superfluid rings.

\section{Homogeneous Systems at Finite Temperature}

%%%%%%%%%%%%%%  FIGURE  %%%%%%%%%%%%%%%%%%%%%%%%%%%%%%%%
\begin{figure}[h]
\begin{center}
   \includegraphics[width=0.45\textwidth,angle=0]{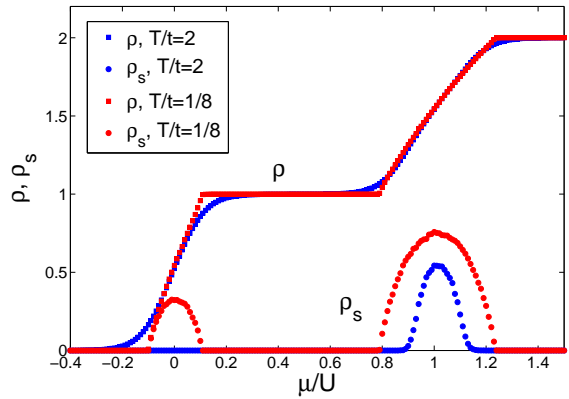}
\end{center}
\vspace{-0.6cm} \caption{\label{fig:rhovsmu} (color online).
Occupation ($\rho$) and SF density ($\rho_s$) as a function of the
chemical potential ($\mu/U$) for two different temperatures
$T/t=2$ and $T/t=1/8$ at $t/U=0.025$ for the homogeneous 2D
Bose-Hubbard model calculated using QMC simulations on a $32
\times 32$ lattice. A nonzero $\rho_s$ corresponds to a superfluid
(SF) phase, zero $\rho_s$ with integer occupation $\rho=1$ and
vanishing compressibility to a Mott insulator (MI) phase, and zero
$\rho_s$ with non-integer occupation to a normal (N) phase. The
key point is that at the low temperature, wherever the density is
incommensurate, the system is also SF. There are only MI and SF
regions, and no N liquid. On the other hand, at higher
temperature, normal regions develop which have incommensurate
filling but also $\rho_s=0$. This is made clear in the figure at
the incommensurate density region near $\mu/U=0$ where a rise in
temperature from $T/t=1/8$ to $T/t=2$ has destroyed the superfluid
density turning the SF to N. Looking at the $\rho_s$ values, it is
also evident that the superfluid properties at higher occupation
($1 < \rho < 2$) such as around $\mu/U=1$ is stronger than that of
the low occupation ($0 < \rho < 1$) around $\mu/U=0$.}
\end{figure}
%%%%%%%%%%%%%%%%%%%%%%%%%%%%%%%%%%%%%%%%%%%%%%%%%%%%%%%%

%%%%%%%%%%%%%%  FIGURE  %%%%%%%%%%%%%%%%%%%%%%%%%%%%%%%%
\begin{figure}[ht]
\begin{center}
   \includegraphics[width=0.5\textwidth,angle=0]{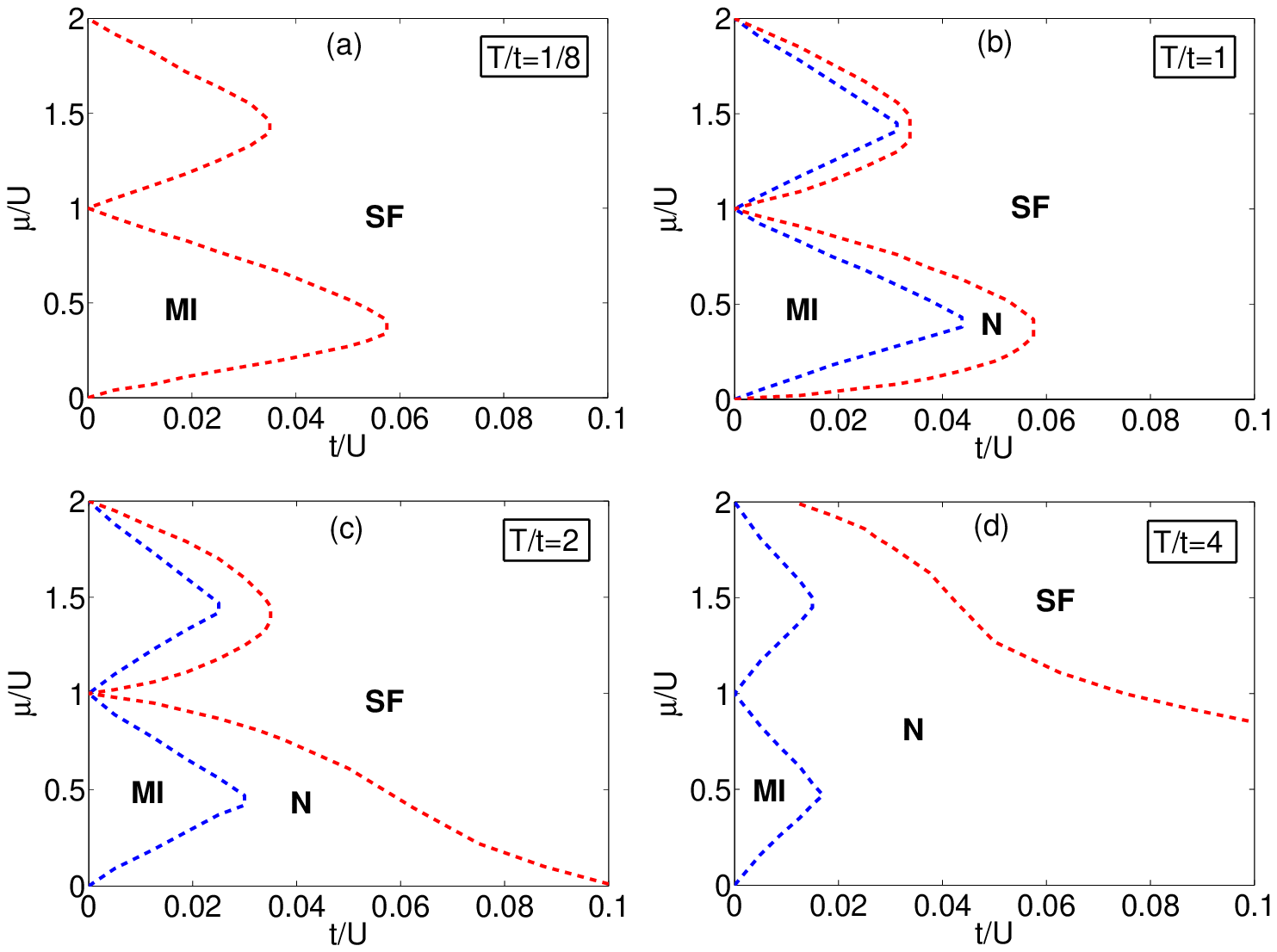}
\end{center}
\vspace{-1.1cm}
\begin{center}
   \includegraphics[width=0.5\textwidth,angle=0]{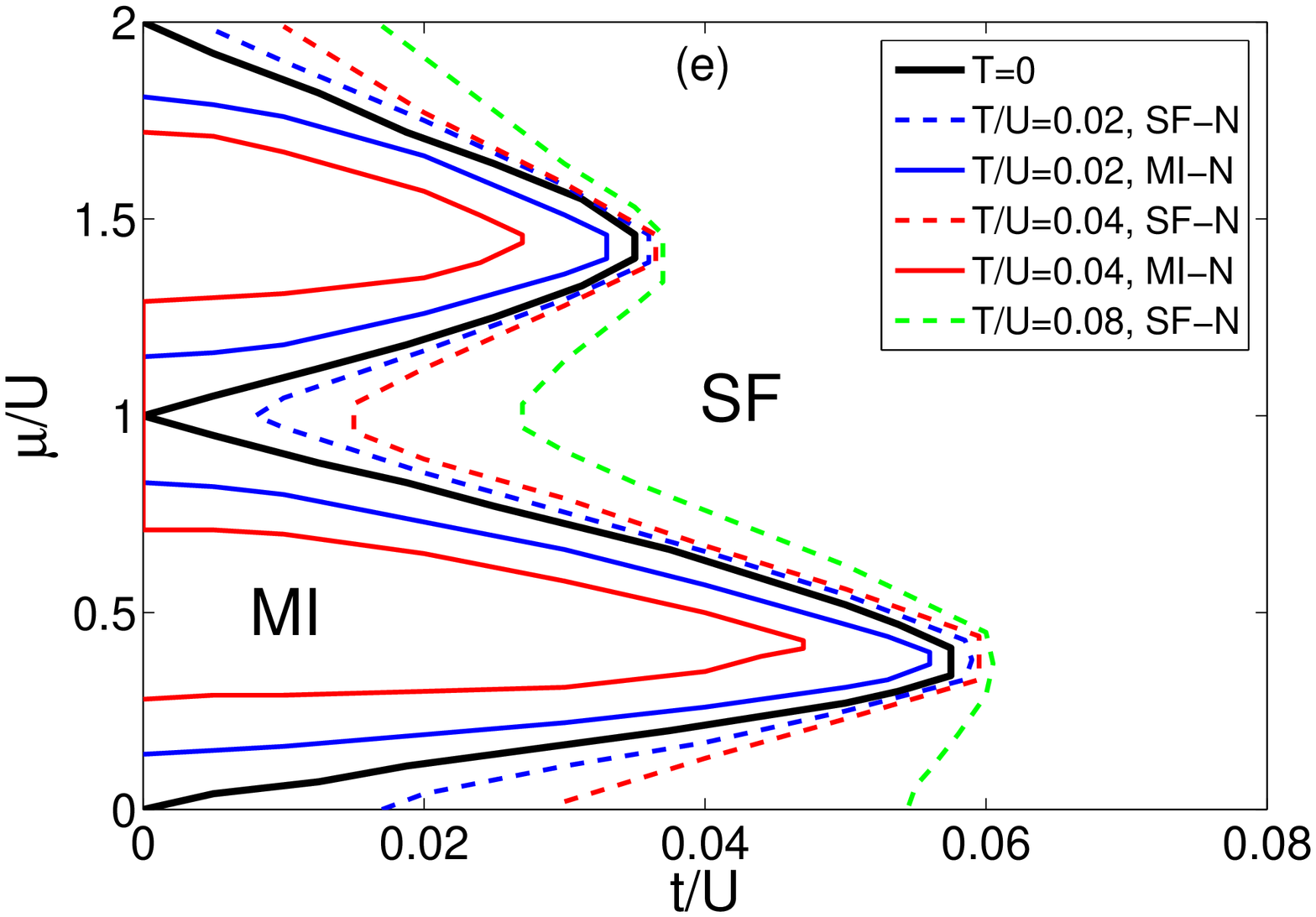}
\end{center}
\vspace{-0.6cm} \caption{\label{fig:homogPhase} (color online).
Finite temperature phase diagram for the homogeneous Bose-Hubbard
model in two dimensions in the ($\mu/U, t/U$) plane for four
different temperatures (a) $T/t=1/8$, (b) $T/t=1$, (c) $T/t=2$ and
(d) $T/t=4$. (e) shows the phase diagram in a different
representation, for constant $T/U$ values of 0.02, 0.04 and 0.08.
The lines that demarcate the SF and N is a phase boundary, and the
lines that demarcate MI and N is a crossover. At finite
temperature, normal phase regions appear between MI and SF.  These
normal regions grow bigger for higher temperature. Although the
finite-T homogeneous phase diagram is usually presented as in
panel (e), with $T$ in units of $U$, the phase diagram plots
(a)-(d) for constant $T/t$ values connect well to the confined
system state diagram presented in Sec. IV, and provide a useful
representation for optical lattice experiments.}
\end{figure}
%%%%%%%%%%%%%%%%%%%%%%%%%%%%%%%%%%%%%%%%%%%%%%%%%%%%%%%%

The zero temperature ground state phase diagram for the
homogeneous Bose-Hubbard model ($V_T=0$) in two dimensions has
been worked out using strong coupling expansions
\cite{freericks96}, and most recently using QMC \cite{sansone08}.
In Ref.~\onlinecite{sansone08}, a finite temperature phase diagram for a
constant occupation $\rho=1$ was presented showing the boundaries
of the SF and MI/N regions in the ($T/t,U/t$) plane. Mean-field
theory has been used \cite{sheshadri93,lu06,bounsante04} to study
different properties and to obtain the finite-T phase diagram in
any dimension. Various other approximate methods have also been
used to study the finite temperature Bose-Hubbard
model~\cite{polak09,ohliger08}. We present here the QMC
homogeneous finite-T phase diagram in the ($\mu/U,t/U$) plane
valid for the first two commensurate filling lobes.

In Fig.~\ref{fig:rhovsmu} the density $\rho$ and the SF density
$\rho_s$ are shown as functions of $\mu/U$ for two different
temperatures $T/t=1/8$ and $T/t=2$ for a $32 \times 32$ lattice at
$t/U=0.025$. (This lattice size is large enough so that finite
size effects are minor.) For $T/t=1/8$, $\rho_s$ is non-zero
whenever $\rho$ is incommensurate.  At this low temperature the
system is either a MI or a SF.  On  the other hand, for $T/t=2$,
this is no longer the case. There are substantial regions where
$\rho_s$ is zero even though $\rho$ is incommensurate, which
signifies a N liquid. For a homogeneous system, the boundary
between the MI and normal regions in the finite temperature phase
diagram is a crossover, as opposed to being a true phase
transition at the SF-N boundary. The MI phase strictly speaking
exists only at $T=0$. However, Mott-like features can persist at a
finite temperature with a small value of compressibility. As
mentioned in Sec. II, in drawing this MI-N crossover boundary at
finite-T, we have taken $\kappa < 0.02/U$ to identify the Mott
region.

The QMC homogeneous system phase diagram for the 2D Bose-Hubbard
model at finite temperature can be built up by sets of runs such
as shown in Fig.~\ref{fig:rhovsmu} for different $U/t$, and is
presented in Fig.~\ref{fig:homogPhase}. We show the phase
boundaries for four different temperatures in panels (a)-(d);
$T/t=1/8$, which is a low enough temperature to be the zero
temperature phase diagram, $T/t=1$, $T/t=2$ and $T/t=4$. At
$T/t=1/8$ the system has only two phases -- MI with an energy gap
inside the lobe, and a gapless SF outside the lobe. The transition
between them is driven by quantum fluctuations as $t/U$ is varied.
The lowest value of $U/t$ for which the SF-MI transition can occur
is $U/t \approx 16.74$ (i.e. $t/U \approx
0.059$)~\cite{sansone08}, the location being at the tip of the
first lobe. At a finite temperature, thermal fluctuations play a
part, and MI lobes are reduced. For example, at $T/t=1$ in
Fig.~\ref{fig:homogPhase}(b), we see that between the MI and SF
phases, a region of normal phase has developed at the expense of
reducing both the SF and MI regions. With further increase of
temperature, the normal phase region widens. For $T/t=2$, the N-SF
phase boundary around $\rho=1$ is no longer lobe-like: the SF
region ceases to bend back inward to low $t/U$ at small density.
With further increase of temperature to $T/t=4$, the SF between
the first two MI lobes is also replaced by N liquid.

In Fig.~\ref{fig:homogPhase}(e) we present the finite-T phase
diagram for temperatures in units of $U$, for $T/U=0.02, 0.04$ and
$0.08$. As temperature increases, the MI lobes shrink, and the N
liquid phase region grows between the MI and SF region. For
$T/U=0.08$ we find that the compressibility $\kappa > 0.02/U$ near
$t/U=0$, which is therefore no longer a MI in our criteria. The
pictorial representations of how the MI lobes behave, although
different in the two representations of $T/t$ and $T/U$, are
interchangeable. For a constant $T/t$ phase diagram, as $t/U$
decreases to zero, $T/U$ also goes to zero, and the MI lobes in
(a)-(d) for low $t/U$ slowly approaches $\mu/U=0,1$ and $2$,
giving rise to a lobe that looks pointy. Although finite-T
homogeneous phase diagram is usually presented at constant $T/U$
as in panel (e), phase diagram plots (a)-(d) for constant $T/t$
values connect well to the confined system state diagram presented
in Sec. IV, and provide a useful representation for optical
lattice experiments where temperatures are often
cited~\cite{spielman10,spielman0708} in units of $T/t$.

We conclude this section by noting that Fig.~\ref{fig:homogPhase}
can be reinterpreted as giving the local density and SF profiles
in a confined system, assuming the validity of the LDA.  That is,
to each spatial site in a confined lattice are associated the
density and SF density of a homogeneous system with the same local
chemical potential.  The resulting sequence of LDA phases which
occurs in moving away from the trap center can most easily be
understood by starting with the location $(\mu/U,t/U)$, where
$\mu$ is the chemical potential at the trap center, on the
homogeneous phase diagram of Fig.~\ref{fig:homogPhase} and moving
downwards, since decreasing $\mu/U$ corresponds to increasing
$V_{T} \, r_i^2$ and thus moving outward from the trap center.

%%%%%%%%%%%%%  FIGURE  %%%%%%%%%%%%%%%%%%%%%%%%%%%%%%%%
\begin{figure}[ht]
\begin{center}
   \includegraphics[width=0.5\textwidth,angle=0]{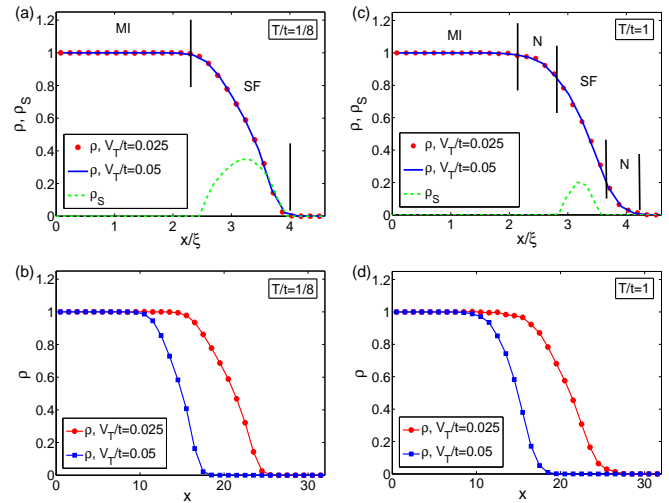}
\end{center}
\vspace{-0.6cm} \caption{\label{fig:density} (color online). The
validity of the characteristic density idea and the effects of
temperature on spatially separated phases in a harmonically
trapped system are illustrated. For $T/t=1/8$, panel (a) shows the
density profiles for two systems with the same characteristic
density $\tilde \rho=34.5$ but different trapping potentials and
total number of particles, $V_T/t=0.025$, $N_b=1380$ and
$V_T/t=0.05$, $N_b=690$. All position dependent quantities are
equivalent\cite{rigol09} as a function of characteristic length
scale $\xi=\sqrt{t/V_T}$, as evident here in density matching. (b)
shows the actual density profiles for the two trapping strengths,
highlighting how these two different spatial profiles match in the
scaled plot (a). Density profiles in (a) exhibit a constant
integer plateau and an incommensurate ring with $n<1$, calculated
using QMC for a trapped system. (c) shows the density profile at a
higher temperature, $T/t=1$, also showing a similar constant
plateau and an incommensurate ring. To determine whether the ring
contains a SF or N liquid, we plot LDA derived SF density across
the trap which shows that for (a) the entire ring is SF, whereas
in (c) only a portion of the ring is SF and the rest is N: $\rho$
is incommensurate but $\rho_s=0$. (d) shows the unscaled density
profiles for the two trapping strengths in (c).}
\end{figure}
%%%%%%%%%%%%%%%%%%%%%%%%%%%%%%%%%%%%%%%%%%%%%%%%%%%%%%%%

\section{State diagram at finite temperature}

%%%%%%%%%%%%%  FIGURE  %%%%%%%%%%%%%%%%%%%%%%%%%%%%%%%%
\begin{figure}[ht]
\begin{center}
   \includegraphics[width=0.4\textwidth,angle=0]{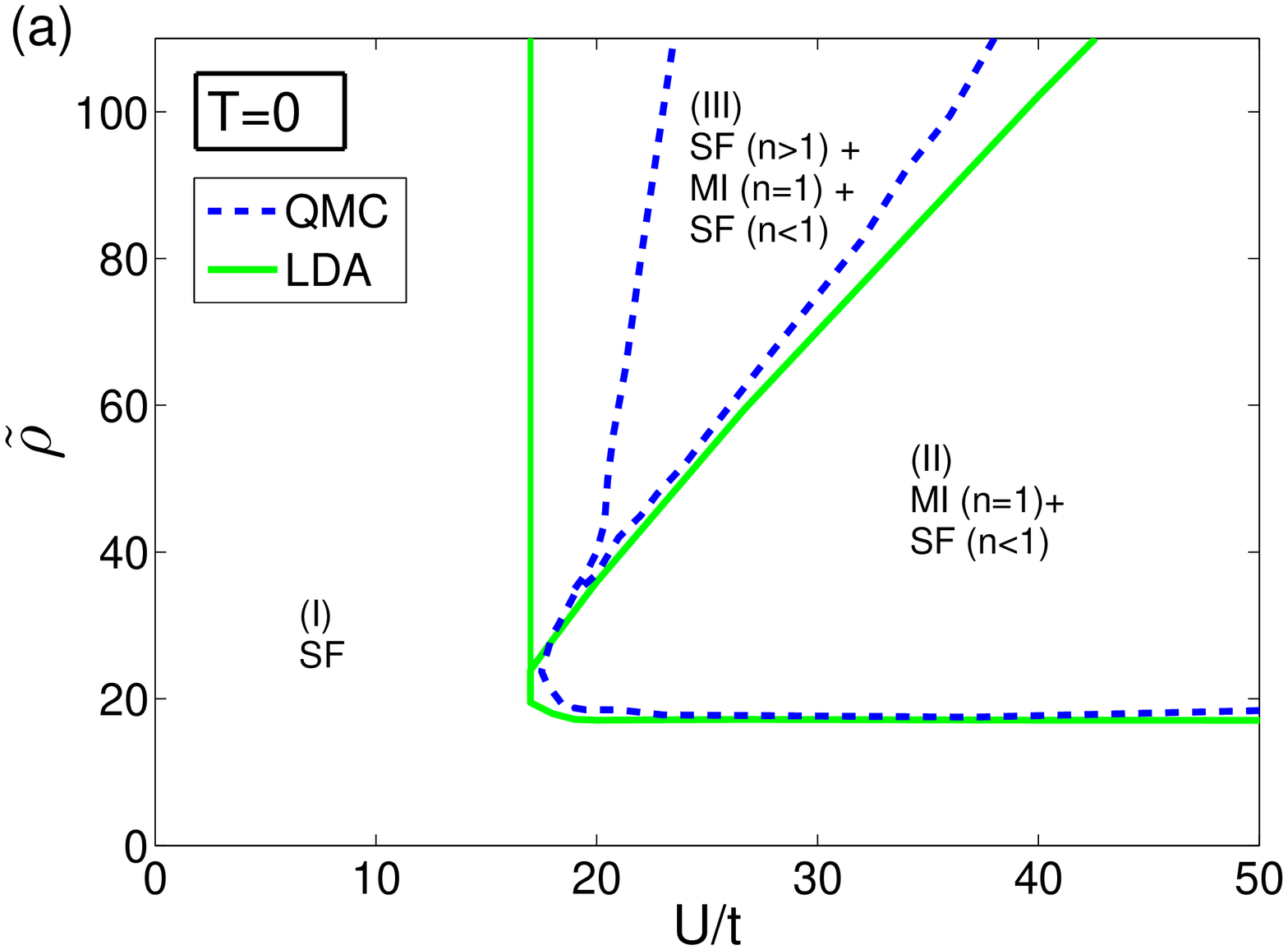}
\end{center}
\vspace{-0.8cm}
\begin{center}
   \includegraphics[width=0.4\textwidth,angle=0]{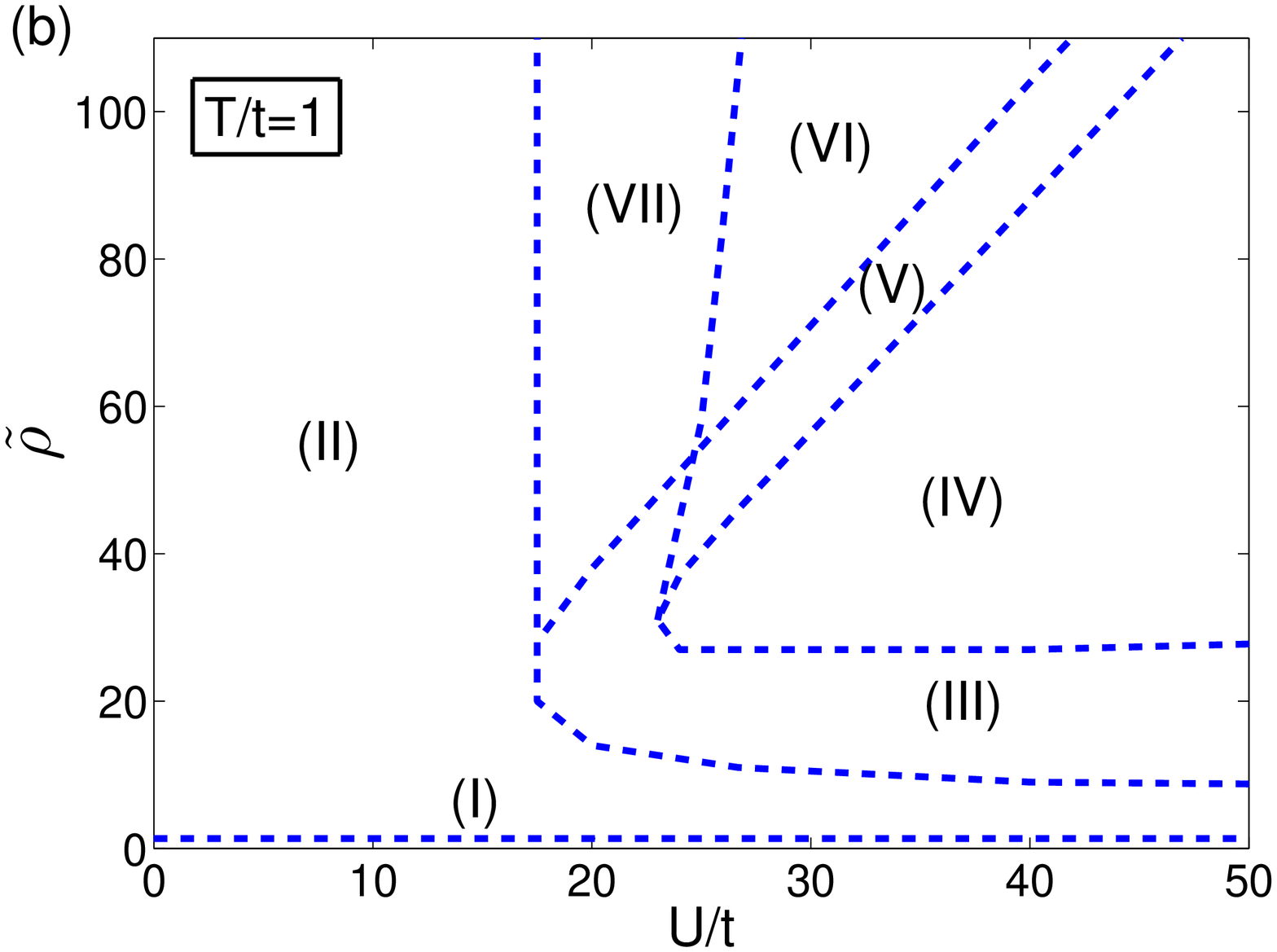}
\end{center}
\vspace{-0.8cm}
\begin{center}
   \includegraphics[width=0.4\textwidth,angle=0]{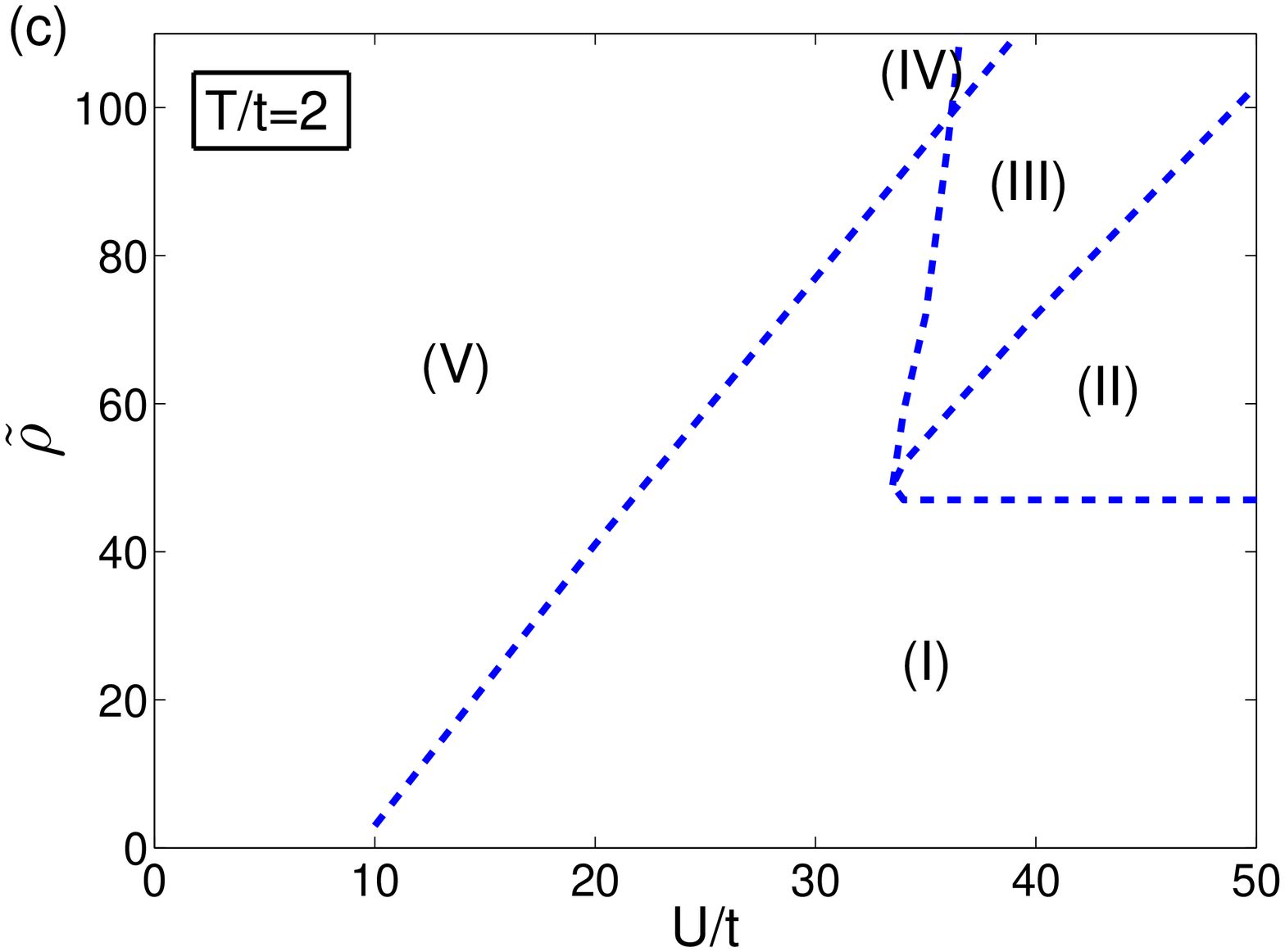}
\end{center}
\vspace{-0.8cm} \caption{\label{fig:trappedPhase} (color online).
State diagrams for a harmonically trapped system for (a) $T=0$,
(b) $T/t=1$ and (c) $T/t=2$. A schematic of all the phases
according to their labels is shown in
Fig.~\ref{fig:phaseSchematic}. The phase boundaries are drawn
according to the multiple phases that coexist in a region. For
example, in (a), region (II) corresponds to a state which has a MI
plateau in the center surrounded by a SF shell. For higher T,
normal (N) phase appears and therefore there are states with many
varieties of coexisting phases. For example, phase (IV) in (b) has
a MI in the center, surrounded outward by N, SF and N
respectively. At a finite temperature, the tail always contains a
normal liquid phase. For higher T, the SF-N phase boundary shifts
to a lower value of $U/t$, and the first appearance of a MI
plateau occurs at a higher value of $U/t$.}
\end{figure}
%%%%%%%%%%%%%%%%%%%%%%%%%%%%%%%%%%%%%%%%%%%%%%%%%%%%%%%%

%%%%%%%%%%%%%  FIGURE  %%%%%%%%%%%%%%%%%%%%%%%%%%%%%%%%
\begin{figure*}[ht]
\includegraphics[width=0.9\textwidth,angle=0]{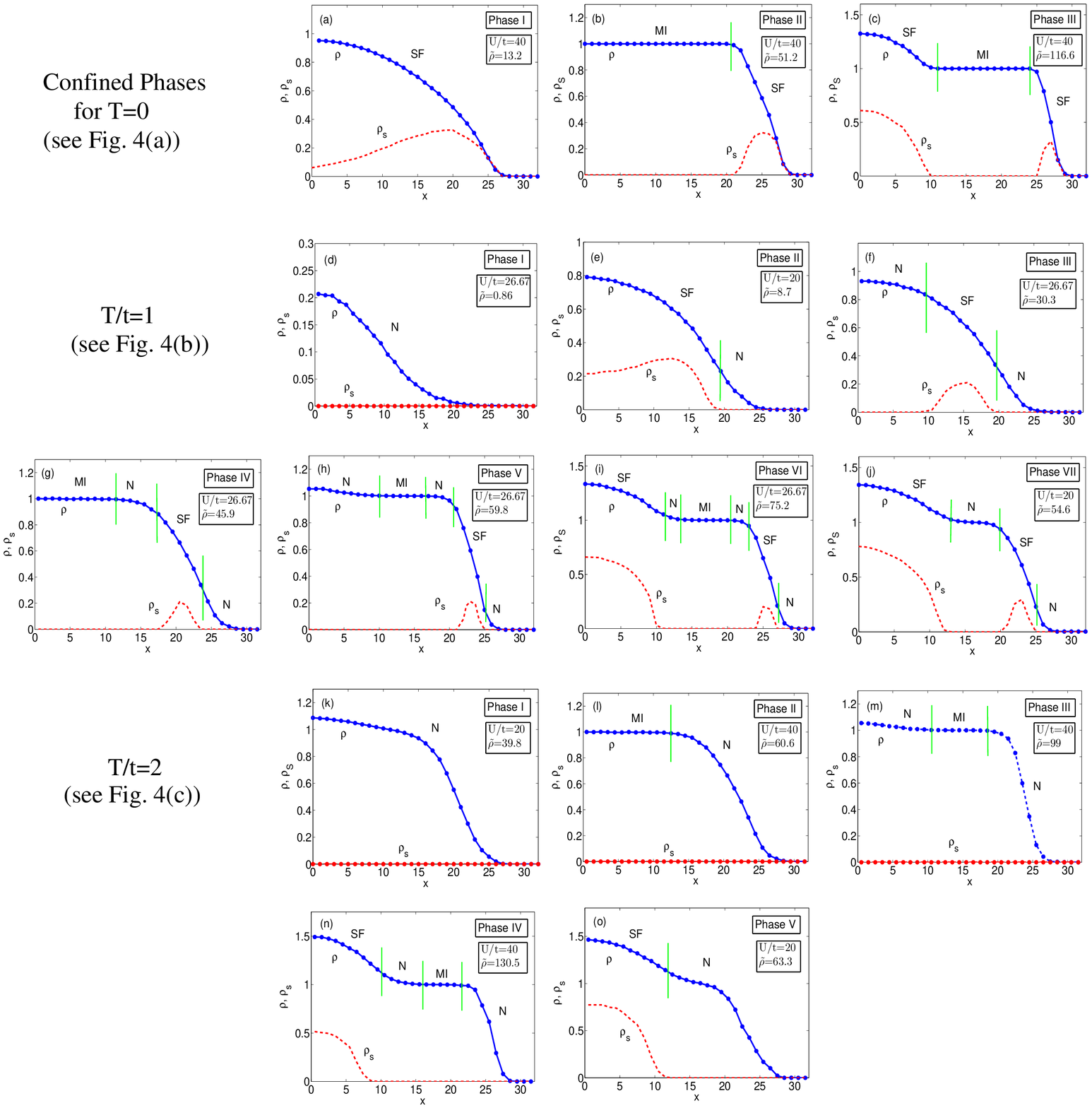}
\vspace{-0.2cm} \caption{\label{fig:phaseSchematic} (color
online). Schematic of all the phases for a trapped system at
$T=0$, $T/t=1$, and $T/t=2$, according to the labels shown in the
state diagrams in Fig.~\ref{fig:trappedPhase}. At finite-T, part
of the atomic cloud turns normal forming rings surrounding SF and
MI regions, and three phases at $T=0$ increase to seven different
phases at $T/t=1$ and five at $T/t=2$. The furthest part of the
tail is always N at a finite-T as seen here in panels (d)--(o).
Annular SF rings cannot form for a high enough temperature such as
at $T/t=2$ shown in panels (k)--(o).}
\end{figure*}
%%%%%%%%%%%%%%%%%%%%%%%%%%%%%%%%%%%%%%%%%%%%%%%%%%%%%%%%

\subsection{Characteristic density}
For the Bose-Hubbard model in the homogeneous case ($V_T=0$), the
thermodynamic limit is reached by increasing the linear lattice
size $L$ while keeping the density $\rho=N_b/L^d$ constant, where
$N_b$ is the particle number, and $d$ is the dimensionality. The
phase diagram is a function only of $U/t$, $T/t$ and $\rho$, not
of $N_b$ and $L$ separately. In the same way, for the harmonically
confined inhomogeneous system with $V_T \neq 0$, the
generalization of this procedure~\cite{rigol09} is to define a
rescaled length, $\xi_i \equiv x_i/\xi$ with $\xi = \sqrt{t/V_T}$,
and a `characteristic density', $\tilde \rho = N_b/\xi^d$. For the
2D model, $\tilde \rho = N_b V_T/t$. The state diagram and the
properties of the inhomogeneous system then depend only on the
combination $\tilde \rho$. Measurements at fixed $U/t$, $T/t$ and
$\tilde \rho$ then approach a well-defined large $L$ limit.
Position dependent quantities match when plotted in units of
rescaled length ($x_i/\xi$). Any LDA-derived quantities like
$\rho$ or local $\rho_s$ trivially follow the rescaling.

The validity of characteristic density idea discussed in
Ref.~\onlinecite{rigol09} is illustrated in
Fig.~\ref{fig:density}. For $T/t=1/8$ (low enough to be
effectively zero temperature), (a) shows the QMC density profiles
for two systems with the same characteristic density $\tilde
\rho=34.5$ but different trapping potentials and total number of
particles, $V_T/t=0.025$, $N_b=1380$ and $V_T/t=0.05$, $N_b=690$.
When plotted as a function of characteristic length scale
$\xi=\sqrt{t/V_T}$, the density profiles are equivalent, as are
all the position dependent local quantities~\cite{rigol09}. (b)
shows the actual density profiles for the two trap strengths in
(a), clearly showing the different extent of the profiles and
highlighting the concept of equivalence in the scaled plots in
(a). The same holds true at finite-T, as shown in
Figs.~\ref{fig:density}(c) and (d) for $T/t=1$. The density
profile in (a) shows a constant integer plateau and an
incommensurate ring with $n<1$, calculated using QMC for a trapped
system. In panel (c) the density profile at a higher temperature,
$T/t=1$, also shows a similar constant plateau and an
incommensurate ring. To determine whether the ring contains a SF
or N liquid phase, we plot LDA-derived SF density across the trap.
This shows that for Fig.~\ref{fig:density}(a) the entire
incommensurate ring is SF ($\rho_s \neq 0$), whereas in
Fig.~\ref{fig:density}(c) only a small part of the ring is SF and
the rest is N, where $\rho$ is incommensurate but $\rho_s=0$. This
is how we identify trapped phases containing distinct combinations
of MI, SF and N regions. Due to the validity of characteristic
density scaling, the state diagram we obtain below is
scale-invariant -- not depending on the specific values of trap
strength ($V_T$), number of particles ($N_b$) or lattice size, but
on the combination $N_b V_T/t=\tilde \rho$. To obtain the state
diagrams presented here, the largest lattice we used is $64 \times
64$.

\subsection{Zero temperature state diagram}
The state diagram for the harmonically confined system at zero
temperature was presented in Ref.~\onlinecite{rigol09}. In
Fig.~\ref{fig:trappedPhase}(a), we show the $T=0$ state diagram in
characteristic density -- interaction strength ($\tilde \rho$,
$U/t$) plane using QMC simulations in a trap, to higher values of
$\tilde \rho$ than presented in Ref.~\onlinecite{rigol09}. Three
separate states are possible here as indicated in
Fig.~\ref{fig:trappedPhase}(a), and illustrated in
Fig.~\ref{fig:phaseSchematic}. The region (I) corresponds to SF
phase all across the trap. In region (II), there is a MI plateau
with ($n=1$) in the center, surrounded by a SF ring of $n < 1$.
(III) is a state (see Fig.~\ref{fig:phaseSchematic}(c)) where
there is a local SF region at the center with $n > 1$, surrounded
by a MI domain with $n=1$ which is further surrounded by a SF ring
with $n<1$. This state diagram quantifies the parameter regimes
for the appearance of these coexistent phases. Knowing the trap
strength, total number of particles, and interaction strength we
can predict the phase in a trap. A MI is obtained for a $U/t$ that
is always greater than the homogeneous system critical coupling of
$(U/t)_{\rm c} \approx 16.74$. Only for a small window of $\tilde
\rho \approx 23$, are the critical couplings comparable. A recent
determination of the critical coupling by a NIST
group~\cite{spielman0708} can be understood in terms of the
characteristic density trajectory the experiment follows when
$U/t$ ratio is increased~\cite{rigol09}. Further
experiments~\cite{spielman10} validate the agreement with the QMC
trapped system state diagram.

To go beyond the results of zero temperature state diagram in
Ref.~\onlinecite{rigol09}, we compare $T=0$ QMC state diagram with
a state diagram obtained by $T=0$ LDA. The green line in
Fig.~\ref{fig:trappedPhase}(a) is the LDA state diagram obtained
from the homogeneous phase diagram by evaluating the density
profiles and characteristic densities for specific values of
chemical potential at the phase boundaries~\cite{batrouni08}. The
obvious disagreement is in the boundary between phase I (SF) and
phase III (SF+MI+SF) which arises due to the finite gradients in
the QMC simulations (and the experiments). Specifically, the
disagreement between phases (I) and (III) boundary is related to
the appearance of finite width MI shoulders. In the LDA picture
the MI appears whenever we are at a chemical potential directly
above the tip of the lobe in Fig.~\ref{fig:homogPhase}. On the
other hand, in a confined system, the shoulder develops more
slowly as we move above $(U/t)_{\rm c}$ and only for a shoulder of
sufficient width does the confined MI compressibility equal that
of a homogeneous system with the same $U/t$ and integer
occupation. The QMC boundary therefore slowly deviates from the
straight line LDA boundary upwards. A recent 2D experiment done at
NIST~\cite{spielman10} has observed this deviation from the LDA.
The transition from phase (II) to phase (III) requires the
development of a superfluid bulge at the center of the trap.  This
occurs when it is favorable to put a particle at the center rather
than at the outer edge of the atomic cloud. By an energy matching
condition, it is possible to show that $\tilde \rho=\pi U/t$,
which implies that the slope of boundary (II)-(III) should be
$\pi$. This is indeed the case for both the LDA and trapped QMC
data points. For MI shoulders with $n \geq 2$ at higher $\tilde
\rho$, similar analytic phase boundaries exist with slopes $5
\pi$, $9 \pi$, etc.

\subsection{Finite temperature state diagram}
Figs.~\ref{fig:trappedPhase}(b) and (c) show the finite
temperature state diagrams at $T/t=1$ and $T/t=2$ respectively.
The phase boundaries demarcate the distinct states that are
possible with different parts of the atomic cloud being a
superfluid, normal fluid or Mott insulator.
Fig.~\ref{fig:phaseSchematic} illustrates all the states of
Figs.~\ref{fig:trappedPhase} (b) and (c) according to their
labels. In order to better understand all the phases, we can start
at a low $\tilde \rho$ at constant $U/t=40$, and think what would
happen if we put more particles in the trap, thereby increasing
$\tilde \rho$ along a vertical line at $U/t=40$ in
Fig.~\ref{fig:trappedPhase}(b). For low $\tilde \rho<1.35$, N
phase extends all across the trap (phase I). Putting more
particles introduces a SF region at the center of the trap while
the tail stays normal (phase II). Putting further particles in the
trap introduces N liquid at the center while the SF+N regions of
the previous phase remain, getting smaller (phase III). Adding
more particles introduces Mott plateau at the center, with the
N+SF+N region getting pushed outside (phase IV). Following the
phase schematics of Fig.~\ref{fig:phaseSchematic} makes this
process clear that adding more particles introduces a new phase
region at the center while keeping the old phases, pushing them
out, and reducing the extent of the regions. This process explains
further appearance of phases V and VI with additional N and SF
regions at the center. We can also think of varying $U/t$ while
keeping a constant $\tilde \rho=60$. For $U/t=10$, we start with a
SF+N phase we encountered earlier. With increasing $U/t$ we reach
phase VII where the center is a SF surrounded by N+SF+N.

With further increase of temperature to $T/t=2$ the state diagram
in Fig.~\ref{fig:trappedPhase}(c) now has five phases as shown in
Fig.~\ref{fig:phaseSchematic}. By focusing on a constant $U/t=40$,
and increasing characteristic density (by putting more particles
in a trap of fixed strength), we can visualize going through the
phases, for this state diagram as well. First we encounter phase I
with N liquid all across the trap. Further increase in number of
particles introduces a MI at the center (phase II). Similarly, in
phases III and IV, N and SF regions appear at the center pushing
out the existing regions. If we start with N phase all across the
trap (phase I) at a lower interaction $U/t=20$ instead of
$U/t=40$, increasing $\tilde \rho$ would introduce a SF bulge at
the center (phase V). Note that $T/t=2$ is high enough in
temperature so that no confined SF rings can appear in the trap,
as is the case for $T=0$ and $T/t=1$.

Some observations of finite temperature effects in a harmonically
trapped lattice system are summarized below: (A) Thermal
fluctuations always introduce a N phase region in the furthest
tail of the atomic cloud where the density is small. As
temperature increases, there always appear N rings surrounding
both MI and SF regions separately, thus giving rise to more than
three confined phases -- seven at $T/t=1$ and five at $T/t=2$.
Thermal effects prevent SF rings to form for high enough
temperature such as we have seen in going from $T/t=1$ to $T/t=2$.
(B) Finite temperature causes a shift of the SF-N transition
coupling $(U/t)_{\rm c}$ to a lower value than for a zero
temperature trapped system, but to a higher value for the first
appearance of a MI plateau. To see this, let us examine the zero
temperature state diagram in Fig.~\ref{fig:trappedPhase}(a) where
the phase boundary between states I and III separates the
appearance of MI shoulders. As we raise the temperature to $T/t=1$
such as in Fig.~\ref{fig:trappedPhase}(b), this boundary (SF-MI)
moves to bigger $U/t$ between the states VII and VI. While another
boundary (SF-N) appears between states II and VII at lower $U/t$
that signifies the appearance of N phase in the center of the
cloud. For temperature range $T=0$ to $T/t<1$ as in NIST
experiment~\cite{spielman10}, the separation between these two
boundaries is small, and the MI transition would pass through a
narrow range in $U/t$ going through a SF-N transition first.  In a
$(T/t-U/t)$ phase diagram at constant integer ($n=1$), this
lowering of $U/t$ for SF-N transition can be understood in terms
of a downward shift of temperature as has been observed and
explained in a recent experiment~\cite{trotzky09} in 3D. (C) In
this paper, we identify and classify the trapped phases with QMC
simulations, and do not study the issue of their experimental
signatures. However, we would like to make some comments in
relation to the state diagrams. There are two main ways of
determining critical coupling in experiments -- by analyzing
momentum distribution (condensate fraction or peak width) in
time-of-flight (TOF) experiments~\cite{spielman0708} and by direct
(in situ) imaging of density profiles~\cite{cchin09}. For zero
temperature trapped system, the critical coupling $(U/t)_{\rm c}$
for SF-MI transition correspond to the appearance of a MI shoulder
observed through a decrease in condensate fraction. At finite-T,
the difference between coherence and incoherence as measured
through condensate fraction differentiates between SF and N/MI.
The distinction between N and MI may be difficult to detect with
TOF imaging. For identifying local phases through the measurement
of a global quantity, a larger portion (than the mere presence of
N/MI) of the atoms has to turn N/MI to detect the signature of its
initial appearance. Since at a finite-T in a harmonic trap, there
is a part of the cloud (in the tails) that is always normal and
part of the cloud (annular rings) that is not a 2D superfluid as
discussed in section V, signatures of the SF-N transition critical
coupling $(U/t)_{\rm c}$ in the momentum distribution of a
harmonically trapped system need to be carefully studied. In situ
imaging may be a better method at finite-T to detect the
appearance of normal phase and MI plateau. Single site resolution
microscopes~\cite{bakr09,sherson10} is a big step forward in this
direction; detection of coexisting phases through density scanning
has been discussed in Ref~\onlinecite{zhou09}.

\subsection{Comparison with NIST experiment}

%%%%%%%%%%%%%  FIGURE  %%%%%%%%%%%%%%%%%%%%%%%%%%%%%%%%
\begin{figure}[ht]
\begin{center}
   \includegraphics[width=0.45\textwidth,angle=0]{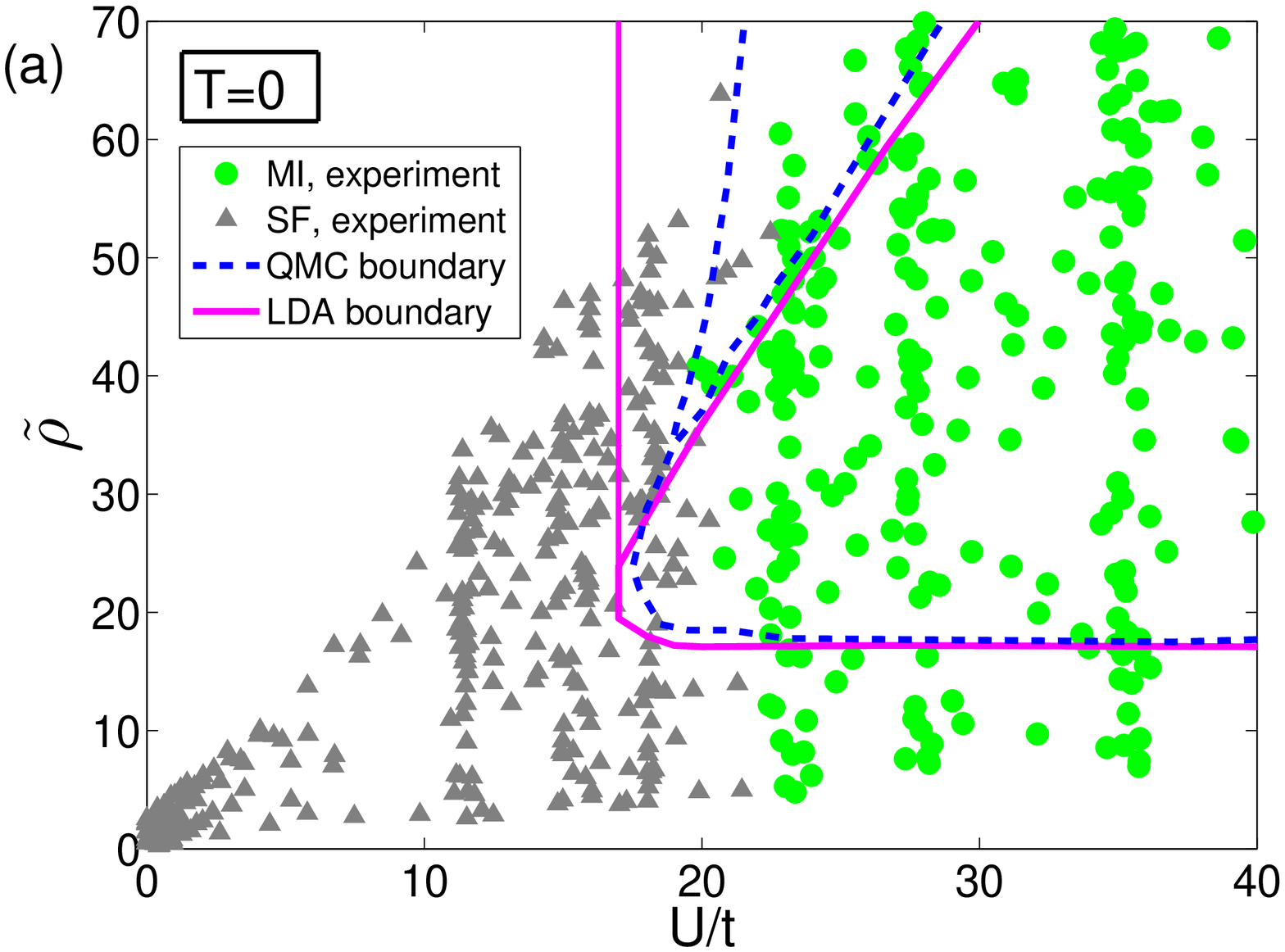}
\end{center}
\vspace{-0.8cm}
\begin{center}
   \includegraphics[width=0.45\textwidth,angle=0]{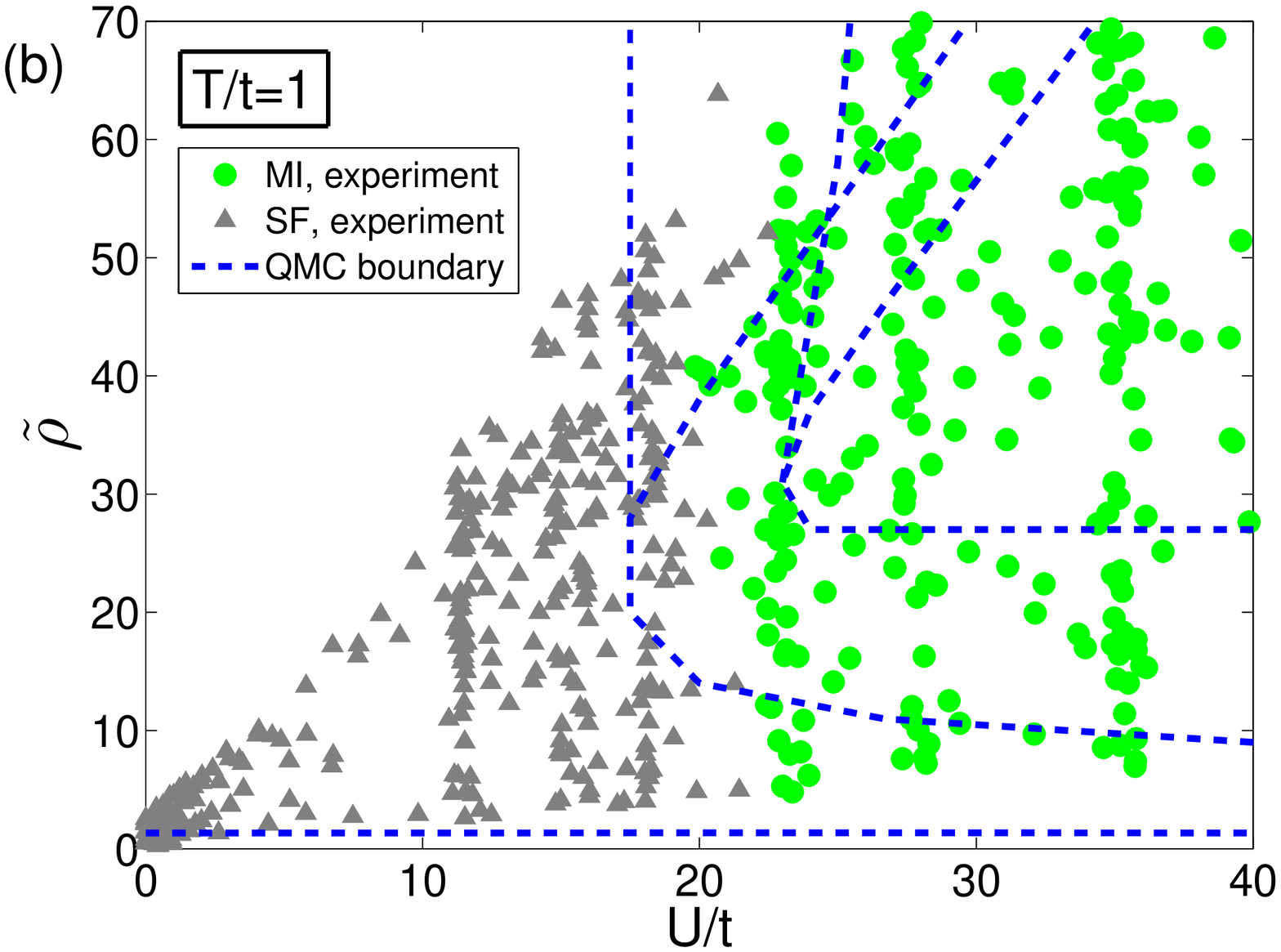}
\end{center}
\vspace{-0.6cm} \caption{\label{fig:nistexp} (color online).
Comparison of theoretical state diagram boundaries with NIST
experimental results of Ref.~\onlinecite{spielman10}. Circles and
triangles represent respectively MI and SF data points from the
NIST experiment; same set of data is plotted in (a) and (b).
Comparison is made with theoretical state diagram boundaries in
(a) $T=0$ and (b) $T/t=1$. The types of coexisting phases for (a)
and (b) are given in Fig.~\ref{fig:trappedPhase}. There are MI
points for low values of $\tilde \rho$ in (a) that are below the
state diagram boundary $\tilde \rho \approx 18$. For finite-T, the
state diagram boundary goes down as in (b) containing many of
these MI/N data points. We identify this feature as a finite
temperature effect fully explainable with a finite-T state
diagram.}
\end{figure}
%%%%%%%%%%%%%%%%%%%%%%%%%%%%%%%%%%%%%%%%%%%%%%%%%%%%%%%%

In a recent experiment at NIST~\cite{spielman10}, the state
diagram for a 2D harmonically trapped lattice system was obtained
by measuring the condensate fraction to identify the superfluid
and Mott insulator regions as a function of atom density and
lattice depth. By comparing to the $T=0$ QMC state diagram, they
have shown the breakdown of LDA, indicated by the appearance of MI
shoulders for higher values of $(U/t)_{\rm c}$ than a homogeneous
system; a phenomenon illustrated in our theory study here in
Fig.~\ref{fig:trappedPhase}(a).

In Figs.~\ref{fig:nistexp}(a) and (b) we show respectively the
$T=0$ and $T/t=1$ state diagrams of
Figs.~\ref{fig:trappedPhase}(a) and (b) overlaid with the
experimental data points of the NIST experiment; the circles
represent MI and the triangles SF. We do not show the experimental
phase boundary and its uncertainty reported in
Ref.~\onlinecite{spielman10}. Although the experimental comparison
was done with a $T=0$ state diagram, the temperature was reported
to be $T/t=0.9(2)$, considered a low enough temperature for such a
comparison. However, as we have seen in
Figs.~\ref{fig:trappedPhase}(a) and (b), because of the appearance
of normal phase regions in the cloud, the state diagram at $T/t=1$
begin to differ from $T=0$. Here we choose to overlay the
experimental data both in our $T=0$ and $T/t=1$ state diagrams
(same data in both panels in Fig.~\ref{fig:nistexp}). It is
visually quite clear that the experiment obtains SF points for low
$U/t$ and MI points for high $U/t$. However, the transition occurs
at different values of $U/t$ for different $\tilde \rho$ in the
vertical axis, shown more clearly in the phase boundary plotted in
Ref.~\onlinecite{spielman10}, signifying breakdown of LDA.

We find that for MI points in low $\tilde \rho$, the data agrees
more with the state diagrams at $T/t=1$ than at $T=0$. For low
$\tilde \rho$ the measurement of different phases is difficult due
to increased thermal fluctuations, and the measurement uncertainty
is large. In Fig.~\ref{fig:nistexp}(a), there appear many MI
points below the boundary of $\tilde \rho \approx 18$. In
Fig.~\ref{fig:nistexp}(b), we see that the lower boundary for MI/N
goes down to $\tilde \rho \approx 9$ containing many of these
points. We should emphasize that the reported experiment
temperature of $T/t=0.9$ is the initial temperature, while the
final temperature during the measurement is unknown and could be
higher, and in such a case, the boundary would go further down
containing all the experimental MI/N data points. With the results
of MI points below the lower boundary for $T=0$, we would argue
that the experiment was able to detect a finite temperature
effect, fully explainable with a finite-T state diagram. We would
like to point out that with time of flight imaging of momentum
distribution and measuring of condensate fraction, the distinction
that is made in the experiment between SF and MI at $T=0$ is the
same as the distinction between SF and N/MI at finite temperature.

\subsection{Role of LDA-derived superfluid density in the state diagram}

%%%%%%%%%%%%%  FIGURE  %%%%%%%%%%%%%%%%%%%%%%%%%%%%%%%%
\begin{figure}[ht]
\begin{center}
   \includegraphics[width=0.4\textwidth,angle=0]{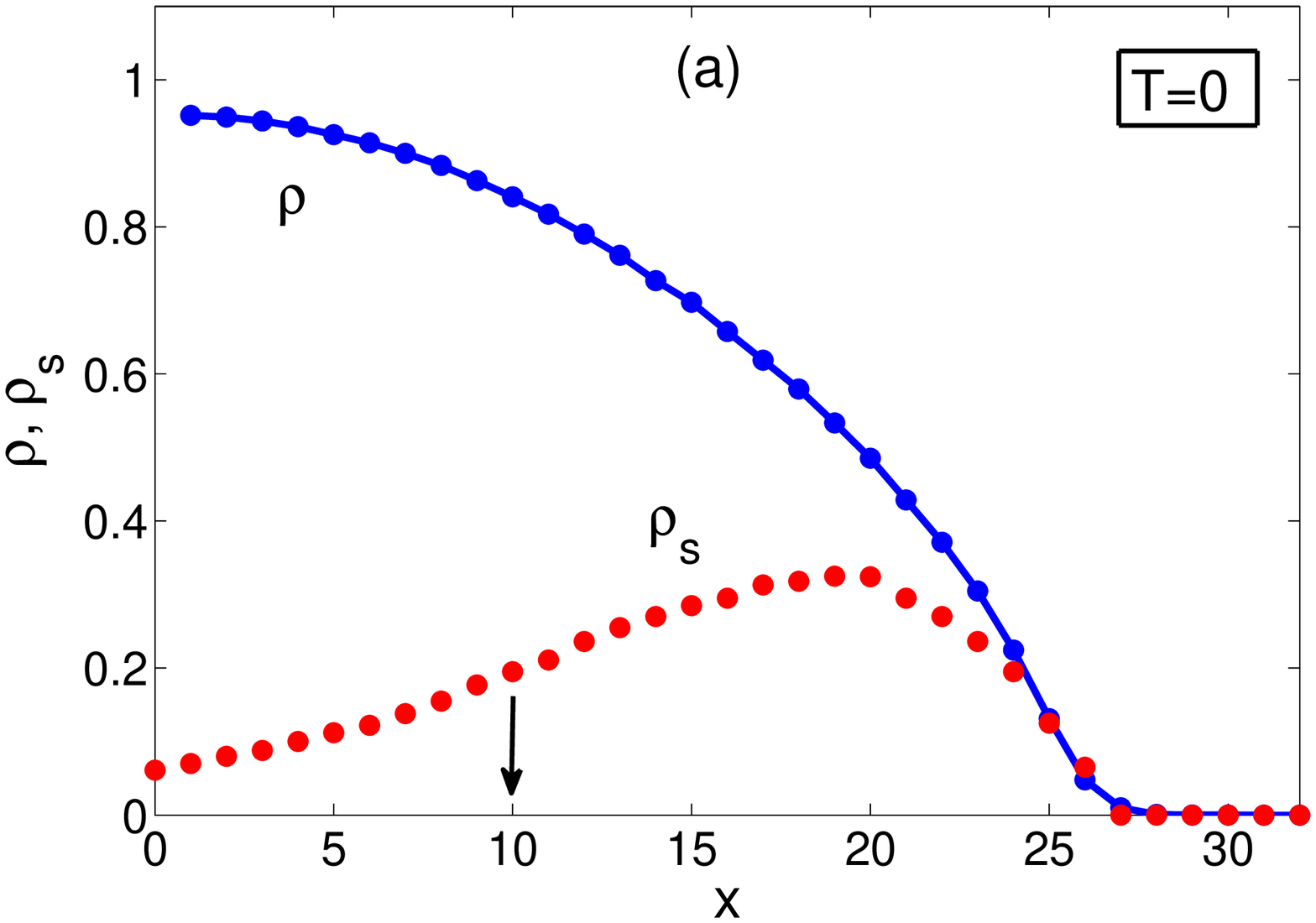}
\end{center}
\vspace{-0.9cm}
\begin{center}
   \includegraphics[width=0.4\textwidth,angle=0]{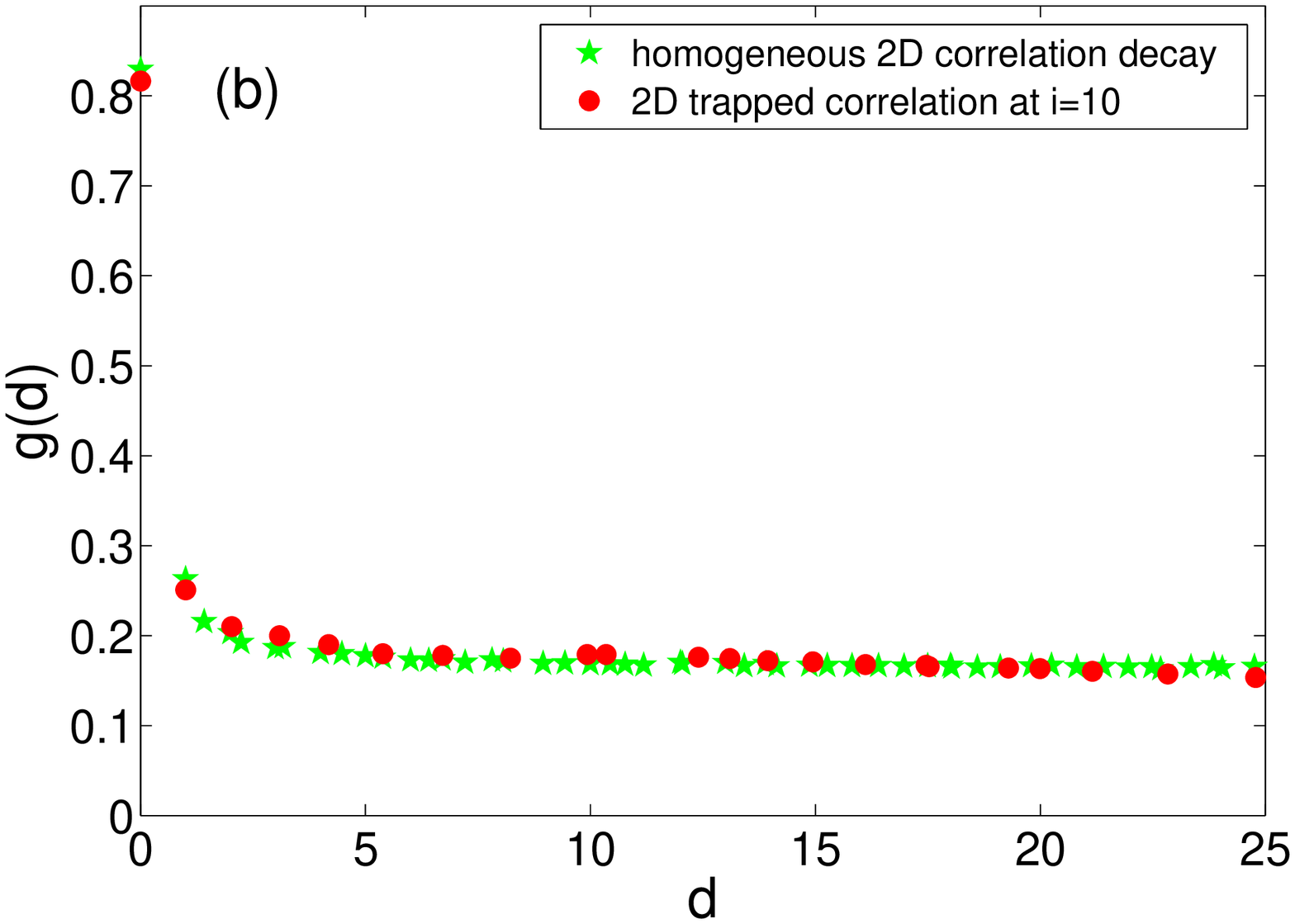}
\end{center}
\vspace{-0.6cm} \caption{\label{fig:sftrap} (color online). (a)
Density profile and local superfluid density for trapped atoms
when the whole region is SF for parameters $t/U=0.025$, $T/t=1/8$,
$V_T/t=0.012$ and $\mu_{center}/U=0.1$. In (b) circles show the
correlation decay between a point $i=10$ and other points along
the circumference of radius $i=10$, as a function of the arc
distance $d$. The pentagrams show correlation decay for a 2D
uniform system for a chemical potential same as that in the
trapped system at $i=10$. Since they match very well, we can
conclude that the effect of trapping has not changed the long
range correlations in this special case when the SF region is wide
enough.}
\end{figure}
%%%%%%%%%%%%%%%%%%%%%%%%%%%%%%%%%%%%%%%%%%%%%%%%%%%%%%%%

The phase boundaries for finite temperature state diagrams in
Fig.~\ref{fig:trappedPhase}(b) and (c) were obtained determining
the density profile and compressibility with QMC simulations of a
confined system. In addition to these two quantities, and in the
absence of a definition of the local superfluid, LDA-derived
$\rho_s$ was used to determine the trapped superfluid region as
has been assumed in the literature~\cite{zhou09,ho09}. In next
section we study bosonic Green's function which measures the
off-diagonal long range order, and by comparing trapped
correlations with homogeneous (LDA) correlations, we find that LDA
does not correctly capture the trapped superfluid properties. For
SF rings of narrow width, correlation decay do not match a 2D
decay, and therefore assigning local $\rho_s$ values from a 2D
homogeneous system has limited validity. So the state diagram
boundaries we have obtained here give the upper limit boundaries
of confined phases. This is only true for the regions where there
are SF rings. High temperature state diagram such as at $T/t=2$ in
Fig.~\ref{fig:trappedPhase}(c), where no SF rings can form, the
phase boundaries are not influenced by this. In
Figs.~\ref{fig:trappedPhase}(a) and (b), for regions with SF
rings, the superfluid designation with LDA-derived $\rho_s$ has to
be kept in mind. At $T=0$ and $T/t=1$, the SF rings are 1D
superfluids or a crossover between 1D and 2D.

To be clear, there can be two LDA issues; one is due to finite
size, and another one is in the description of superfluid
properties in trapped rings with LDA $\rho_s$, an effect we
identify in next section. Since QMC simulations in a trap were
used for density and compressibility, we do not have the
finite-size LDA issue related to the appearance of MI shoulders
like in $T=0$ diagram in Fig.~\ref{fig:trappedPhase}(a). It is the
use of local LDA $\rho_s$ to designate the phases, for which we
would like to qualify our state diagrams by the fact that in the
regions where there are annular SF rings, they can exhibit true
long-range order or quasi-long-range order depending on the width
of the ring.

\section{Correlation function at zero and finite temperature}

%%%%%%%%%%%%%  FIGURE  %%%%%%%%%%%%%%%%%%%%%%%%%%%%%%%%
\begin{figure}[ht]
\begin{center}
   \includegraphics[width=0.4\textwidth,angle=0]{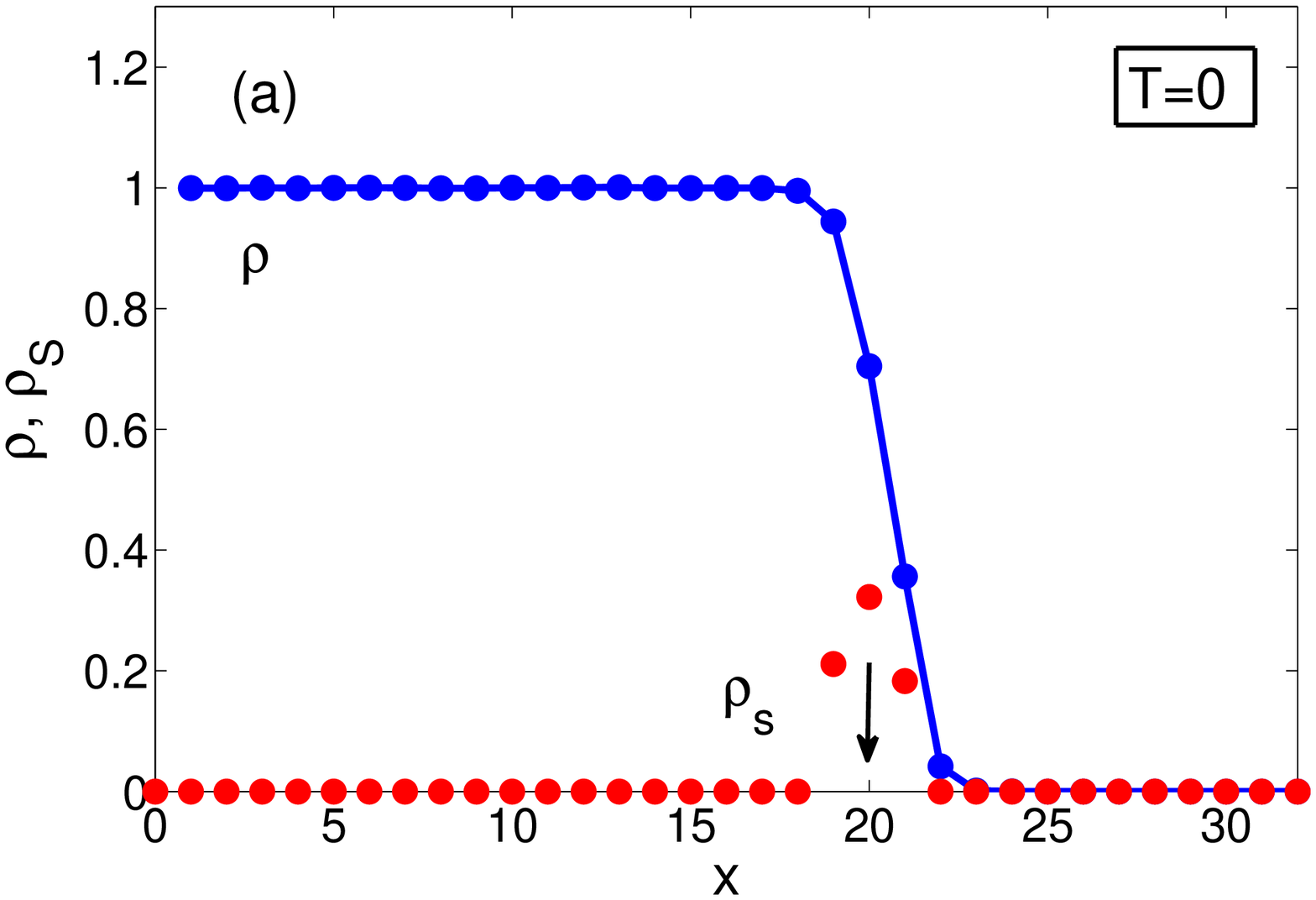}
\end{center}
\vspace{-1.5cm}
\begin{center}
   \includegraphics[width=0.4\textwidth,angle=0]{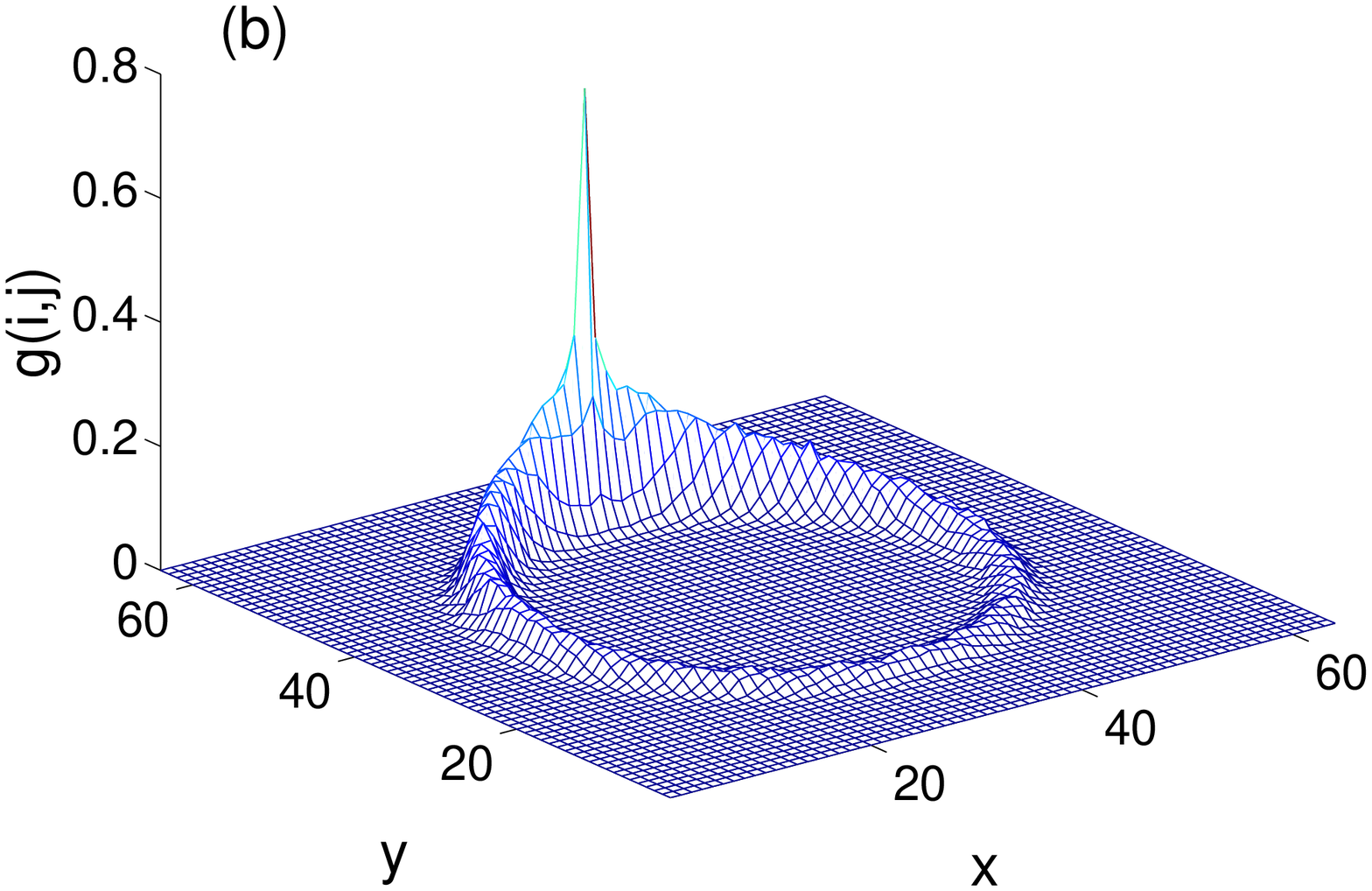}
\end{center}
\vspace{-0.9cm}
\begin{center}
   \includegraphics[width=0.4\textwidth,angle=0]{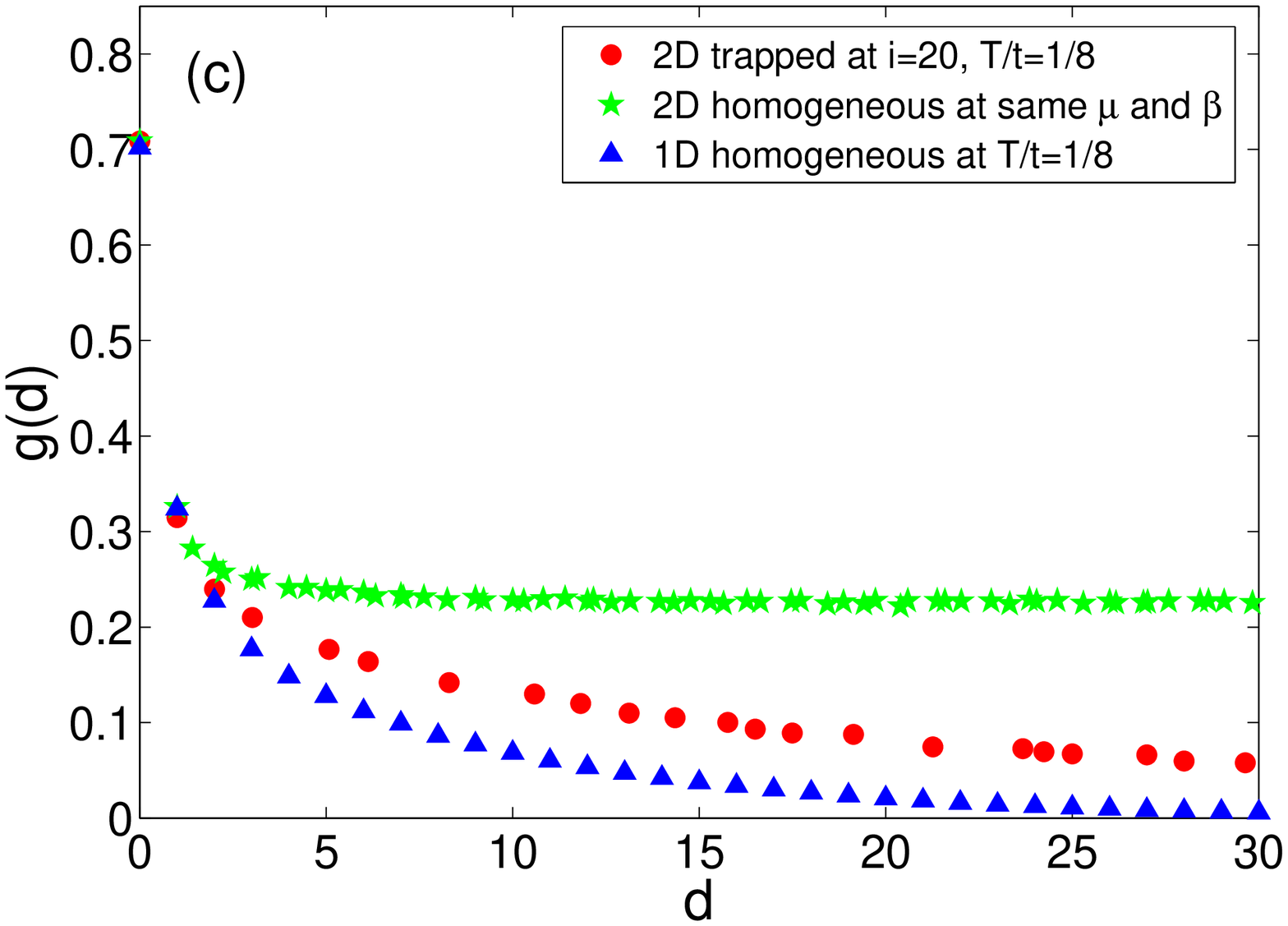}
\end{center}
\vspace{-0.6cm} \caption{\label{fig:narrowRing} (color online).
The effect of reduced dimensionality in a trapped superfluid ring.
Panel (a) depicts a narrow SF ring of only 4 sites in width, for
parameters $t/U=0.025$, $T/t=1/8$, $V_T/t=0.07$ and
$\mu_{center}/U=0.7$. 3D plot of $g(i,j)$ for $i=20$ is shown in
(b), where we see that $g(i,j)$ is nonzero all across the ring.
However, a comparison with the homogeneous correlations in (c)
reveals that the trapped decay (circles) is faster in the ring
than a true 2D decay (pentagrams), matching only for a very small
distance. In (c) we also compare trapped ring correlations with a
1D correlation decay (triangles) at the same parameters and
temperature $T/t=1/8$, and find that the 1D algebraic SF decay is
faster. This demonstrates that the bose gas in the incommensurate
ring is in a crossover regime between 1D and 2D. Assigning 2D LDA
$\rho_s$ values to the trapped rings therefore has limited
validity, pointing out a breakdown of LDA in describing superfluid
properties in a trap.}
\end{figure}
%%%%%%%%%%%%%%%%%%%%%%%%%%%%%%%%%%%%%%%%%%%%%%%%%%%%%%%%

%%%%%%%%%%%%%  FIGURE  %%%%%%%%%%%%%%%%%%%%%%%%%%%%%%%%
\begin{figure*}[ht]
\includegraphics[width=0.9\textwidth,angle=0]{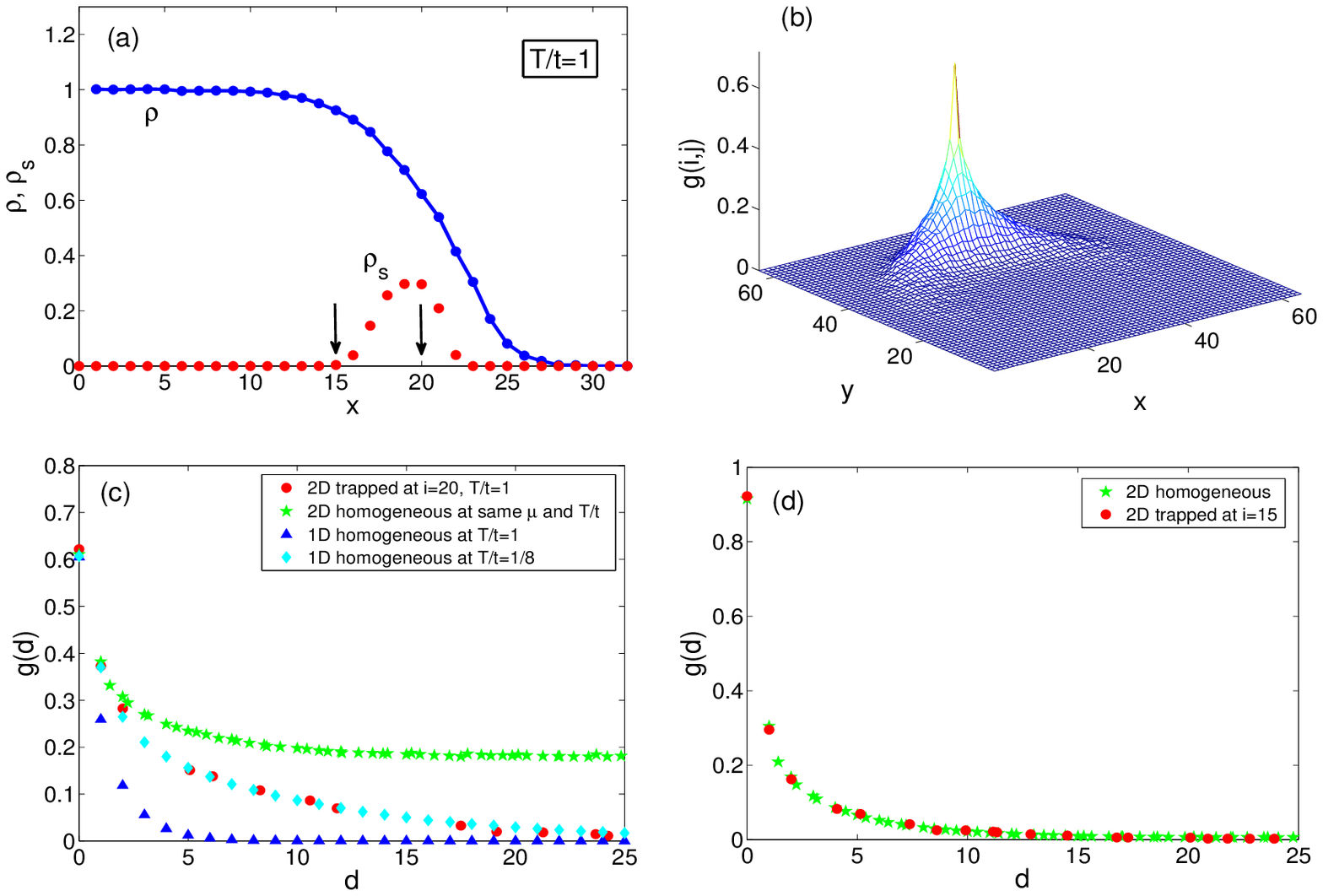}
\vspace{-0.6cm} \caption{\label{fig:finiteTgij} (color online).
The behavior of spatial correlations for trapped superfluids at
finite temperature. For the parameters $T/t=1$, $\mu/U=0.4$,
$t/U=0.05$ and $V_T/t=0.4$, (a) shows the density profile and
local superfluid density indicating a superfluid ring of finite
width. (b) plots $g(i,j)$ between the point $i=20$ and the rest of
the lattice, showing clearly that although $i=20$ is SF according
to LDA picture, the decay is such that there is no long-range
order along the entire ring. (c) quantifies the decay in greater
depth -- the circles showing the correlation along the ring slowly
going to zero at a large distance. $g(i,j)$ for a homogeneous 2D
superfluid with the same parameters is given in pentagrams. The
triangles, 1D correlation at the same density and temperature
$T/t=1$, decays much faster. However, the diamonds depicting 1D
correlations at a much lower temperature ($T/t=1/8$) are seen to
be similar to that of the 2D trapped annular SF correlations at
$T/t=1$. This suggests that for the purpose of long range
properties the quasi-1D ring is at a much lower effective
temperature than the equilibrium temperature of the 2D atomic
cloud. For normal region at $i=15$, the 2D homogeneous
correlations match as shown in (d).}
\end{figure*}
%%%%%%%%%%%%%%%%%%%%%%%%%%%%%%%%%%%%%%%%%%%%%%%%%%%%%%%%

To further analyze the trapped system, we study spatial
correlations in the trap by looking at the one-particle density
matrix between bosons on site $i$ and $j$, $g(i,j)=\langle
a^\dagger_{i} a_{j} \rangle$. As shown in previous section, the
harmonic trap at zero and low enough temperature creates a
superfluid ring surrounding the Mott insulating or normal region.
The SF ring, having a finite width and a relatively longer length
around a circle, has a quasi-1D geometry.

For the parameters $\mu/U=0.1$, $t/U=0.025$, $V_T/t=0.012$, and
zero temperature ($T/t=1/8$), Fig.~\ref{fig:sftrap}(a) shows the
density profile of trapped atoms where the whole region is SF. The
red circles in (a) are local superfluid densities obtained by LDA.
In Fig.~\ref{fig:sftrap}(b), we plot the trapped correlation
function decay (circles) between a point at $i=10$ and all points
along the ring at the same radius, as a function of their distance
$d$. Since the circular ring has the same local chemical potential
at all points, we can take a homogeneous system with that
specified chemical potential, and make a meaningful comparison
with the trapped correlation function as we do in LDA for
variables such as density, compressibility, etc. Such a
homogeneous correlation decay shown in pentagram symbols match
very well the trapped $g(i,j)$ decay. This implies that the effect
of trapping (inhomogeneity) has not changed the behavior of
long-range correlations; this is true in this specific example
where the SF region is wide enough that it retains its 2D
superfluid properties, as we will soon show that trapping does
have an effect on the SF rings of quasi-1D geometry.

In Fig.~\ref{fig:narrowRing}, we investigate the zero temperature
($T/t=1/8$) correlations for a narrow SF ring. Panel (a) shows the
density profile and local superfluid density indicating that the
ring is approximately 4 sites in width. Panel (b) shows a 3D plot
of $g(i,j)$ for a point in the SF ring, between $i=20$ and for all
other sites $j$ in the lattice. It is evident that order persists
all across the narrow ring as $g(i,j)$ is nonzero along the
circle, whereas it decays rapidly to zero radially from the SF
region to the MI plateau. Similar to Fig.~\ref{fig:sftrap}(b), in
Fig.~\ref{fig:narrowRing}(c) we compare the trapped correlation
decay for $i=20$ along the ring (circles) to the homogeneous
correlations at the same $\mu$ (pentagrams). It shows that the
trapped SF decay matches the homogeneous decay for a short
distance, after which it continues to deviate from a 2D SF decay.
Thus the SF ring does not have 2D superfluid properties, rather it
exhibits quasi-long-ranged correlations, influenced by the width
of the ring. So the trapped annular SF decay not matching a 2D SF
decay, we can find out whether it might match a 1D correlation
decay, since the ring has a finite width. In
Fig.~\ref{fig:narrowRing}(c) we compare 2D trapped correlations
along the ring with 1D homogeneous correlation decay (triangles)
at the same density and temperature, $T/t=1/8$. In 1D, the decay
is algebraic ($|i-j|^{-p}$, $p$ is the decay exponent, a positive
real number) for superfluid and exponential ($e^{-|i-j|/\xi}$,
$\xi$ is correlation length) for a Mott or normal
phase~\cite{kollath04,rigol04,cirac08a}. We see in (c) that the 2D
ring correlations decay slower than a homogeneous 1D algebraic SF
decay but faster than a homogeneous 2D decay. We can therefore
conclude that the bose gas in the SF ring is in a crossover regime
between a 1D and 2D superfluid. As the width narrows, it would
approach a 1D SF decay, and as the width increases, as we have
seen in the earlier example, the trapped correlation decay would
approach a 2D SF decay.

While all these may seem quite intuitive because of the quasi-1D
nature of the ring, the significance of these results is in the
fact that assigning 2D LDA-derived local superfluid density to the
ring, as have been assumed in many recent
papers~\cite{zhou09,fang10} and in previous section, is not
totally justified. It indicates a breakdown of LDA for local
condensate properties. We know that for trapped atomic systems,
LDA describes very well the local density dependent quantities,
such as density, variance and
compressibility~\cite{zhou09,batrouni08}. A known failure of LDA
in trapped systems is due to finite size effects in determining
the appearance of a phase boundary~\cite{rigol09,spielman10}. We
show here that the effect of reduced dimensionality in traps lead
to another failure -- in the LDA description of local superfluid
density in the trapped annular rings.

Fig.~\ref{fig:finiteTgij} illustrates the behavior of spatial
correlations at finite temperature which is one of the key results
of this article. For the parameters $T/t=1$, $\mu/U=0.4$,
$t/U=0.05$ and $V_T/t=0.4$, Fig.~\ref{fig:finiteTgij}(a) shows the
density profile and local superfluid density exhibiting a
superfluid ring of finite width. Fig.~\ref{fig:finiteTgij}(b)
plots $g(i,j)$ between the point $i=20$ and other points on the
lattice, showing clearly that although $i=20$ is SF according to
LDA picture, the decay is such that there is no long-range order
across the entire ring. Fig.~\ref{fig:finiteTgij}(c) quantifies
the decay in greater depth -- the circles showing the correlation
along the ring slowly going to zero at a large distance. For a
homogeneous 2D superfluid with the same $t/U$, $T/t$, and at the
same $\mu$ as the spatial location $i=20$, the correlation is
shown in pentagrams, where the decay is correctly that of a 2D SF.
1D correlation at the same temperature $T/t=1$ and parameters,
shown in triangles, decays much faster. However, the diamonds
depicting 1D correlations at a much lower temperature $T/t=1/8$
are seen to be qualitatively similar to that of 2D trapped annular
ring at $T/t=1$. This implies that as far as the long range
properties are concerned, the annular ring is at a lower effective
temperature than the 2D cloud. This may be due to reduced
fluctuations induced by the collective behavior of a finite width
ring. In Fig.~\ref{fig:finiteTgij}(d) we show correlations in
trapped normal region at $i=15$ where it matches with the 2D
homogeneous correlations.

\section{Conclusion}

In this paper, we determined the combined effects of harmonic
trapping and temperature for a square lattice Bose-Hubbard model
to obtain the finite temperature state diagram.  This extends
previous work\cite{rigol09} which examined phase coexistence at
$T=0$. As temperature increases, thermal fluctuations melt away
both the SF and MI phases, introducing the N phase. At finite-T,
the N liquid phase is always present in a trap, in the lower
density regions furthest from the trap center. Furthermore, each
SF and MI region is surrounded by a N ring. This gives rise to
many different confined phases. As the temperature increases, the
critical coupling $(U/t)_{\rm c}$ for the SF-N transition is
lowered and the MI-N crossover coupling is increased. For the
phases that contain SF annular rings in the state diagram, the
quasi-long range nature of their correlations have to be kept in
mind. In addition to the trapped state diagram, we presented the
homogenous system phase diagram at finite temperature, for
temperature in units of both $T/t$ and $T/U$.

We compare our state diagrams to a recent experiment at
NIST~\cite{spielman10} done on a harmonically trapped 2D lattice.
Although the experiment mainly focused on reporting the breakdown
of LDA in obtaining the critical coupling, we identify that they
have also observed signatures of finite-T in the state diagram.

To further understand the trapped phases, we examine the
dependence of spatial correlations $g(i,j)$ in the annular
superfluid ring. We show that the correlation decay in SF rings
does not match the 2D homogeneous superfluid at the same
parameters. For short distances, on the order of the width of the
ring, the correlations agree, while for longer distances the
deviation gets bigger. At zero temperature, the correlation decay
is intermediate between 1D and 2D decay. At finite temperature,
the trapped correlation decay rate is much faster than a
homogeneous 2D decay. Although it is still slower than a 1D decay
at the same temperature, 1D correlations at a much lower
temperature matches the trapped decay. This indicates that the
ring is at a lower effective temperature, a fact that may have
important consequences for long range properties in lower
dimensions. These studies point out the fact that assigning LDA
$\rho_s$ values to the trapped SF rings has limited validity, and
thus provide evidence for the breakdown of the local density
approximation (LDA) in the description of superfluid properties of
trapped bosons.

The quantitative values for the phase boundaries provided here
provide numerical benchmarks for continuing efforts to emulate the
Bose-Hubbard model on optical lattices, and demonstrate
experimental-theoretical consistency for the numerical values of
the location of the critical points.

\begin{acknowledgments}
This work was supported under ARO Grant No. W911NF0710576 with
funds from DARPA OLE program. KWM acknowledges a travel award from
the Institute of Complex Adaptive Matter (ICAM). We acknowledge
computational support from the Ohio Supercomputer Center. We would
like to thank Karina Jimenez-Garcia and Ian Spielman for providing
their experimental data.
\end{acknowledgments}

{}


\begin{thebibliography}{}
\bibitem{foot1}
To make a dimensionality independent connection, the hopping term $t$
is sometimes written as $t/z$, where $z$ is the
number of nearest neighbor sites, $z=2$ for 1D, $z=4$ for 2D and
$z=6$ for 3D hypercubic lattices \cite{kato08}. Since we are strictly focusing
on 2D, we analyze our results in the convention the
experimentalists use\cite{cchin09,spielman10}
as given in the Hamiltonian in
Eq.~\ref{HubbB}.
\end{thebibliography}
\end{document}